\documentclass[lettersize,journal]{IEEEtran}
\usepackage{amsmath,amsfonts}
\usepackage{siunitx}
\usepackage{algorithmic}
\usepackage{algorithm}
\usepackage{array}
\usepackage[caption=false,font=normalsize,labelfont=sf,textfont=sf]{subfig}
\usepackage{textcomp}
\usepackage{stfloats}
\usepackage{url}
\usepackage{verbatim}
\usepackage{graphicx}
\usepackage{cite}
\usepackage{multirow}
\usepackage{booktabs}
\usepackage{cuted}
\usepackage{ragged2e}
\usepackage{subfig}
\usepackage{gensymb}
\title{3D high-resolution imaging algorithm using 1D MIMO array for autonomous driving application}
\author{Sen Yuan,~\IEEEmembership{Graduate Student Member,~IEEE}, Francesco Fioranelli,~\IEEEmembership{Senior Member,~IEEE},
and Alexander Yarovoy,~\IEEEmembership{Fellow,~IEEE}
\thanks{Copyright (c) 2015 IEEE. Personal use of this material is permitted. However, permission to use this material for any other purposes must be obtained from the IEEE by sending a request to pubs-permissions@ieee.org. 
}
\thanks{The authors are with the Microwave Sensing, Signals and Systems (MS3) Group, Delft University of Technology, Delft, 2628 CD, The Netherlands (e-mail:s.yuan-3@tudelft.nl; f.fioranelli@tudelft.nl; a.yarovoy@tudelft.nl).}% <-this % stops a space
}

\begin{document}
\maketitle

\begin{abstract}
The problem of 3D high-resolution imaging in automotive multiple-input multiple-output (MIMO) side-looking radar using a 1D array is considered. The concept of motion-enhanced snapshots is introduced for generating larger apertures in the azimuth dimension. For the first time, 3D imaging capabilities can be achieved with high angular resolution using a 1D MIMO antenna array, which can alleviate the requirement for large radar systems in autonomous vehicles. The robustness to variations in the vehicle's movement trajectory is also considered and addressed with relevant compensations in the steering vector. The available degrees of freedom as well as the Signal to Noise Ratio (SNR) are shown to increase with the proposed method compared to conventional imaging approaches. The performance of the algorithm has been studied in simulations, and validated with experimental data collected in a realistic driving scenario.

\end{abstract}

\begin{IEEEkeywords}
MIMO array, direction of arrival (DOA) estimation, 3D high resolution imaging, automotive radar.
\end{IEEEkeywords}  

\section{Introduction}
\IEEEPARstart{R}ADAR can provide accurate and direct measurements of the range, relative velocity, and angle of multiple targets, as well as a long-range coverage of over 200 m even in challenging weather or lighting conditions \cite{Advances} outperforming in this sense other sensors, namely, camera and Lidar. Thus, it has attracted significant importance in the intelligent vehicle systems\cite{shishanov2018height,zhu2022measurement,wang2021cfar}. Newly-developed automotive imaging radars provide information in three spatial dimensions (range, azimuth, and elevation), plus the Doppler information, thus generating more accurate azimuth and elevation estimations for 3D imaging, which also contributes to denser point clouds beneficial for later processing \cite{palffy2022multi, prophet2020semantic}. 

The quality of 3D imaging relies on the radar's range and angular resolutions. Frequency-modulated continuous-wave (FMCW) automotive radar working at millimeter wave band can offer a large operational bandwidth, providing sufficient range resolution. Angular resolution in both azimuth and elevation is contingent upon the antenna aperture and thus determined by the number and layout of the transmit and receive antenna elements, limited by the radar cost and packaging size. To provide 3D imaging ability, the antenna array needs to distribute in both the azimuth and elevation dimension. In the context of 3D imaging, multiple-input multiple-output (MIMO) radar can be used to form the so-called virtual array, providing a cost-effective way to improve the angle resolution of the radar. Hence, FMCW MIMO radar is broadly exploited for imaging in autonomous driving \cite{sun2020mimo}.

Performances in terms of achievable resolution and signal-to-noise ratio (SNR) can vary depending on the processing algorithm, such as digital beamforming (DBF) \cite{DBF}, Minimum Variance Distortionless Response (MVDR) \cite{capon_MVDR}, and subspace-based methods, such as MUltiple SIgnal Classification (MUSIC) \cite{xu2018super}, and Estimation of Signal Parameters via Rational Invariance Techniques (ESPRIT) \cite{roy_esprit-estimation_1989}. The ultimate outcome depends significantly on the composition of the virtual array and the number of available channels. A different approach is provided by synthetic aperture radar (SAR) imaging, a technique predominantly used in remote sensing. Essentially, the SAR concept involves exploiting the motion of a platform to synthesize arrays (apertures) of longer, arbitrary length. Inspired by the idea of SAR, automotive radar imaging algorithms uses the vehicle's motion to increase angular resolution. The motion of the platform can be estimated by fusion with other sensors including optical cameras, laser\cite{rasshofer2005automotive}, and conventional navigation sensors such as wheel-based odometry and inertial sensors, or by using radar-based Simultaneous Localization and Mapping (SLAM) algorithms \cite{kellner2014instantaneous}, or other ego-motion estimation algorithms based on the radar raw signal \cite{yuan20233}.

An approach forming a synthetic aperture for automotive MIMO radar has been explored in \cite{gao2021mimo}, which can only enhance the resolution in a limited region. A SAR-inspired method in \cite{iqbal2021imaging} uses back projection to image the whole scene based on targets' estimated trajectories with a high computational cost. The motion-enhanced snapshots were introduced in \cite{yuan2023vehicular} to extend the MIMO aperture coherently, achieving better angular resolutions with a slightly increased computational cost. A super-resolution DOA algorithm based on relative motion was proposed in \cite{zhang2020super}, which can only boost the resolution in a region of interest, i.e., the range of angles where targets have already been detected. Therefore, an additional processing step is needed to first detect the targets and estimate their related DOA values, which is then followed by the process of boosting the angular resolution. The application of Doppler Beam Sharpening (DBS) for the angular resolution refinement of low-Terahertz radar sensing was investigated in \cite{daniel2018application}. Combination of DBS with the fast iterative adaptive approach to achieve high azimuth resolution in the forward-squint region was proposed in \cite{mao2018super}. Implementing interpolations in SAR algorithm for imaging static targets on the sides of the ego-vehicle was proposed in \cite{tebaldini2022quick}. The 'unambiguous Doppler based forward-looking multiple-input multiple-output radar beam sharpening scan' (UDFMBSC) approach was proposed in \cite{yuan2022novel}. It combines MIMO processing with DBS in the signal domain and offers a solution to the ambiguity problem with reasonable computation cost. In subsequent research \cite{yuan2023adaptive}, the 'Robust unambiguous DBS with adaptive threshold' (RUDAT) approach was introduced to account for diverse reflectivity scenarios and movements.

However, all the aforementioned algorithms assume the flat earth model, the elevation angles remain fixed at $\phi=0$, and are typically designed only for 2D imaging, i.e., to generate 2D range-azimuth maps. 3D higher angular resolution algorithms are needed to further improve the radar's imaging ability for better perception of environment. Currently, there are few 3D imaging algorithms available for automotive radar. \cite{xu20223d} addresses the 3D high-resolution imaging problem by introducing a distributed radar system. 3D imaging capability is achieved by a 2D array in the azimuth and elevation dimension, which needs larger array size and relative higher power consumption. 
In this paper, a novel 3D imaging algorithm is proposed to improve imaging capabilities in both azimuth \& elevation compared to the state-of-the-art algorithms, while using just one dimensional MIMO array in the elevation domain, which significantly alleviates the requirement for large radar apertures in vehicle systems. Motion-enhanced snapshots are introduced to generate larger apertures in the azimuth dimension. The robustness to the vehicle's irregular movement is also considered with compensations in the steering vector. With the proposed method, the degrees of freedom increase as well as the SNR. The performance of the proposed method is thoroughly analyzed for ideal point targets and extended targets in simulations, as well as experimental data.

The main contributions of the paper are listed as follows:
\begin{enumerate}
\item{A novel 3D imaging algorithm using a 1D MIMO array is demonstrated, by introducing motion-enhanced snapshots to achieve high angular resolution for side-looking automotive radar. 
}
% \item{The motion-enhanced snapshots are proposed to generate coherent larger aperture in the azimuth dimension, while a new automotive radar installation is using to use the larger virtual aperture formed by MIMO for elevation estimation.}
\item{A formulation of the steering vector is proposed to address the 3D imaging problem jointly in azimuth \& elevation, and to compensate motion artefacts from irregular movement of the ego-vehicle.}
\item{The proposed algorithm is validated with simulated data, as well as experimental data collected by multiple sensors in realistic driving scenarios.}
\item{The use of motion-enhanced snapshots is shown to increase the SNR and the degrees of freedom for DOA estimation, thus enabling to distinguish more targets and solve the rank deficiency problem of MUSIC.}

\end{enumerate}

The rest of this paper is organized as follows. In Section  \uppercase\expandafter{\romannumeral2}, the signal model is provided. The proposed method is described in Section \uppercase\expandafter{\romannumeral3}. The results for simulated point targets, simulated complex extended targets, and experimental data are provided in Section \uppercase\expandafter{\romannumeral4}. A detailed discussion on the proposed method using simulation and numerical analysis is given in \uppercase\expandafter{\romannumeral5}. Finally, Section \uppercase\expandafter{\romannumeral6} concludes the paper.

\section{Signal Model}
\label{section2}
% \subsection{signal model}

\begin{figure} [htbp]
 \centering
 \includegraphics[width=75mm]{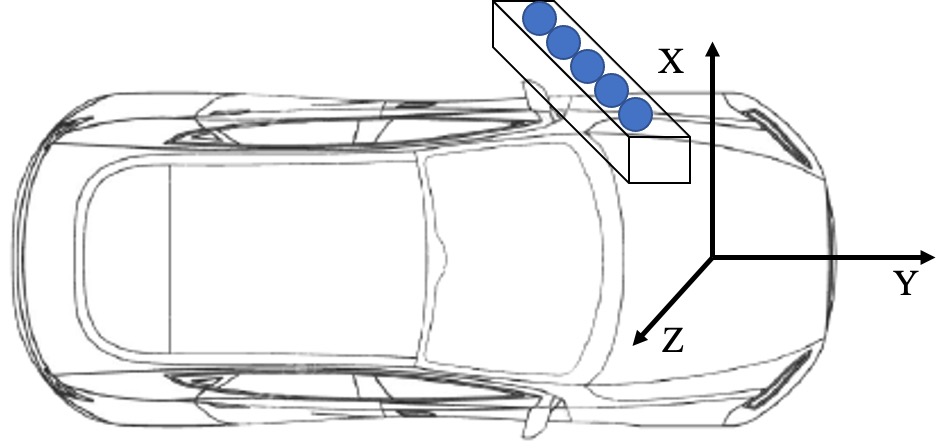}
 \caption{The geometry of the side-looking automotive radar along the Z direction. Y is the forward direction, X is the cross-forward direction, and Z is the elevation direction satisfying the left-handed Cartesian coordinates.}
 \label{geometry}
\end{figure}

FMCW MIMO radar with $N_e$ array elements for elevation estimation and $N_a$ elements for azimuth is considered here. The omnidirectional antenna pattern is considered for the transmitter and receiver, without losing generality. The radar is installed on the side-looking position, where the y-axis refers to the direction of movement of the radar, as shown in Fig. \ref{geometry}.

The radar received signal will be the sum of the reflected signals by the scatter points in the field of view. After reception, the signal is mixed with the transmitted signal to derive the baseband signal (de-chirping). The discretized radar signal will be:

\begin{equation}
\begin{split}
    z(p,q,l,b)&=\sum\limits_{{\mathbf o}}^{k_s}\alpha_{\mathbf o} \exp[j\Phi (\theta_{\mathbf o},\phi_{\mathbf o},p,q)]\\
&\times exp[-j2\pi(f_{0}\frac{2v_{\mathbf o}}{c}T_pl+\mu \frac{\gamma_{\mathbf o} b}{f_s}]\\
\label{fundamental_equation}
\end{split}
\end{equation}
where $z$ indicates the radar received signal; $\mathbf{o}$ is the target's index;  $k_s$ is the total number of scattering points; $p=0,1,2,...,N_a$ and $q=0,1,2,...,N_e$ are the index of the antenna elements counted from the $1$st antenna element in azimuth and elevation direction, respectively; $l=0,1,2,...,L_d-1$ is the slow time index with $L_d$ is the total number of chirps in one frame; $b=0,1,2,...,B_d-1$ is the sampling index in fast time and $B_d={T_cf_s}$ is the maximum number of samples within one chirp; $f_s$ is the sampling frequency; $T_c$ is the chirp duration; $\alpha_{\mathbf o}$ is the constant complex amplitude related to the characteristics of the target ${\mathbf o}$; $f_0$ denotes the starting frequency of the chirp; $c$ is the speed of light; $\theta_{{\mathbf o}}$ and $\theta_{{\mathbf o}}$ are the azimuth and elevation of target ${\mathbf o}$; $v_{{\mathbf o}}$ is the radial velocity between the target ${\mathbf o}$ and radar; $T_p$ is the pulse repetition time (PRT); $\mu$ is the frequency modulation rate; $\gamma _{\mathbf o} =\frac{2D_{{\mathbf o}}}{c} \ll T_c $ with $D_{{\mathbf o}}$ being the distance between antenna and the target ${\mathbf o}$. The term $\Phi$ denotes the phase differences between different antenna pairs.

% \begin{equation}
% \begin{split}
%     z(p,q,l,k)&=\sum\limits_{{\mathbf o}}^{k_s}\alpha_{\mathbf o} \exp[j\Phi (\theta_{\mathbf o},\phi_{\mathbf o},i,j) ]\\
% &\times exp[-j2\pi(f_{0}\frac{2v_{\mathbf o}}{c}T_cl+\mu \frac{\gamma k}{f_s}]\\
% \label{fundamental_equation}
% \end{split}
% \end{equation}
% where $z$ indicates the radar received signal; $\mathbf{o}$ is the target's index; $k_s$ is the total number of scattering points; $i=0,1,2,...,N_e-1$ and $j=0,1,2,...,N_a-1$ are the index of the antenna elements counted from the $1$st antenna element in elevation and azimuth direction, respectively; $l=0,1,2,...,L_d-1$ is the slow time index, with $L_d$ is the total number of chirps in one frame; $k=0,1,2,...,K_d-1$ and $K_d={T_cf_s}$ is the maximum number of samples within one chirp; $f_s$ is the sampling frequency; $T_c$ is the chirp duration; $\alpha_{\mathbf o}$ is the constant complex amplitude related to the characteristics of the target ${\mathbf o}$; $f_0$ denotes the starting frequency of the chirp; $d$ is the space interval between adjacent antenna elements; $c$ is the speed of light; $\theta_{{\mathbf o}}$ is the azimuth of target ${\mathbf o}$; $v_{{\mathbf o}}$ is the radial velocity between target ${\mathbf o}$ and radar; $\mu$ is the frequency modulation rate; $\gamma _{\mathbf o} =\frac{2D_{{\mathbf o}}}{c} \ll T_c $ with $D_{{\mathbf o}}$ being the distance between antenna and target ${\mathbf o}$. The term $\Phi$ denotes the phase differences between different antenna pairs.

\begin{equation}
\begin{split}
    \Phi (\theta_{\mathbf o},\phi_{\mathbf o},p,q) &=\Phi_a (\theta_{\mathbf o},\phi_{\mathbf o},p)+\Phi_e(\phi_{\mathbf o},q) \\
    &\approx 2\pi 
    f_0(\frac{pd}{c} {\rm{sin} \theta_{\mathbf o} \rm{cos} \phi_{\mathbf o}} + \frac{qd}{c} \rm{sin}\phi_{\mathbf o})
\end{split}
\label{angle_information}
\end{equation}
where $d=\frac{\lambda}{2}$ is the space between adjacent antenna elements in MIMO virtual array, $\lambda$ is the wavelength, $\phi_{\mathbf o}$ is the target's elevation angle.

% \subsection{DOA estimation in MIMO}
% \label{DOA}
\section{Proposed Method}
%\subsection{Problem formulation}
In this section, the proposed approach to enable 3D imaging capabilities while keeping the number of antenna elements limited to a 1D array in the elevation direction is described. It is shown how the key step of the approach is the exploitation of the movement of the vehicle by the motion-enhanced snapshots, but limiting the computational complexity and the duration of the accumulated radar data. 
%Automotive radar equipped with 3D high resolution imaging ability will replace other radar as the primary sensor in the autonomous vehicle industry. 
%The number of antenna is always limited because of the power consumption and the size of radar, so there is always the trade-off between the elements used for elevation and azimuth estimation. It is necessary but challenging to provide high angular resolution algorithm for 3D imaging using limited number of antenna.

%The movement information of the radar will become the key parameter, and it is the fundamental information for SAR algorithm, which contributes to the high spatial resolution for SAR processing. Utilizing the motion information can be a big supplement for automotive imaging radar. It has the potential to solve the challenge problem in high-resolution 3D imaging algorithm. Even some research proposed SAR-inspired algorithm in automotive industry, they are computational heaviest and requires accumulating data from a long time, which limits the real-time processing. 

% The concept of using motion information is proposed in \cite{yuan2023vehicular} to solve the DOA estimation with a limited number of snapshots to suit for fast changing scenarios in automotive industry assuming no elevation differences. 

\subsection{Proposed method}

The processing chain of the proposed algorithm includes three blocks, namely the generation of motion-enhanced snapshots, the compensated steering vector, and the 3D beamscan. 

\subsubsection{Motion-enhanced snapshots}
\label{Motion-enhanced snapshots}
As a 1D array in elevation is considered here, the signal in equation (\ref{fundamental_equation}) after range-FFT can be written omitting the dependence on the azimuth index $p$ as:

\begin{equation}
\begin{split}
    \hat{z}(q,l,\hat r) &=\sum\limits_{{\mathbf o}}^{k_s}\alpha_{\mathbf o} \pi T_c f_s \text{sinc} (\frac{(\hat rT_c f_s+\mu \gamma_{\mathbf o}T_c)}{2})\\
    &\times \exp[j( \Phi_e (\phi_{\mathbf o},q) +2\pi f_{0}\frac{2v_{\mathbf o}}{c}Tl)] \\
&\times \exp[-j\pi(mT_cf_s+\mu \gamma_{\mathbf o}T_c)]\\
\end{split}
\label{2d_fft}
\end{equation}
where $\hat r$ is the index of a range bin with detected targets. The second term is related to the antenna elements for spatial and Doppler/phase information. 
% $\Phi$ will become
% Here, since the 1D array in elevation is considered here, $\hat{z}(i,j,l,\hat r)$ will become $\hat{z}(i,l,\hat r)$. $\Phi$ will become
% \begin{equation}
%     \Phi (\theta_{\mathbf o},\phi_{\mathbf o},i,j)=\Phi_e(\phi_{\mathbf o},i) \approx 2\pi 
%     f_0\frac{id}{c} \rm{sin}\phi_{\mathbf o}
% \label{angle_information}
% \end{equation}

If azimuth antenna elements were present, the phase differences between adjacent elements would be: 

\begin{equation}
    \Phi_a(\theta_{\mathbf o},\phi_{\mathbf o},p) \approx 2\pi 
    f_0\frac{pd}{c} \rm{sin}\theta_{\mathbf o} \rm{cos} \phi_{\mathbf o}
\label{angle_information}
\end{equation}
where $\theta_{\mathbf o}$ is the azimuth angle of target $\mathbf o$.

It should be noted that if the slow time index satisfies the below relation in (\ref{motion-enhanced_deriva}), then snapshots at different slow time indices can be used to provide azimuth information coherently. 

\begin{equation}
\begin{split}
  2\pi f_{0}\frac{2v_{\mathbf o}}{c}T(l_1-l_0) =  \Phi_a (\theta_{\mathbf o},&\phi_{\mathbf o},p+1) -\Phi_a (\theta_{\mathbf o},\phi_{\mathbf o},p)\\
\end{split}
\label{motion-enhanced_deriva}
\end{equation}

In side-looking radar, the velocity of the static target will be equal to:
\begin{equation}
\begin{split}
v_{\mathbf o}= v_x \rm{cos} \theta_{\mathbf o} \rm{cos} \phi_{\mathbf o} +v_y \rm{sin} \theta_{\mathbf o} \rm{cos} \phi_{\mathbf o} +v_z \rm{sin}\phi_{\mathbf o}
\end{split}
\label{Doppler_velocity}
\end{equation}

As the major component of the vehicle's motion in the considered geometry will be the velocity in the Y direction, equation (\ref{motion-enhanced_deriva}) can be rewritten and approximated as follows, where the approximation is necessary in this stage and will be compensated in the later processing.

\begin{equation}
\begin{split}
 \frac{d}{c} \rm{sin} \theta_{\mathbf o} \rm{cos} \phi_{\mathbf o}&= \frac{2v_y \rm{sin} \theta_{\mathbf o} \rm{cos} \phi_{\mathbf o} }{c}T(l_1-l_0 )\\
\Rightarrow l_1 &=l_0+\lfloor{\frac {d}{2v_yT}}\rfloor T
\end{split}
\label{snapshot_formation}
\end{equation}

To satisfy the constraint of an integer slow time index, the $\lfloor \rfloor$ rounding operation is implemented in the equation (\ref{snapshot_formation}), thus introducing an approximation error.

The motion-enhanced snapshots for the selected antenna vector at time $l_0$ will be defined as the group of snapshots whose slow time index follows the following equation:

\begin{equation}
l_n=l_0+n T_{ind}, i \in \mathbb{Z}
\label{new definition of small aperture}
\end{equation}
where $T_{ind}=\lfloor{\frac {d}{2v_yT}}\rfloor T$ is the approximate time tag interval for a coherent aperture extension. 

\begin{figure}[h]
 \centering
 \includegraphics[width=65mm]{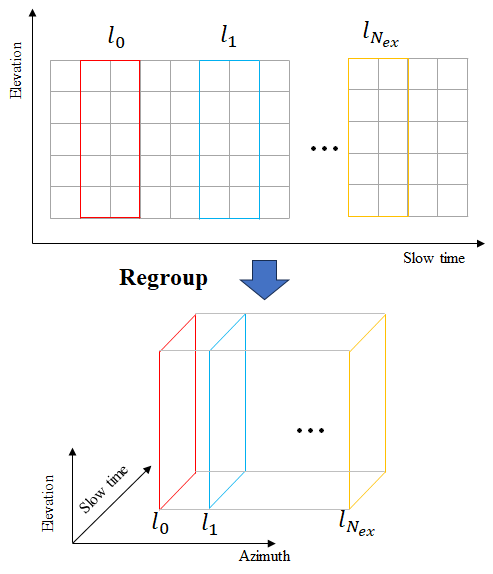}
 \caption{The reshaping process to form a 3D data cube in elevation, azimuth, and slow-time domain from 2D data. }
 \label{regroup}
\end{figure}

The radar signal can be reshaped according to the time tag as shown in Fig. \ref{regroup}. As this reshaping operation of the entire 4D tensor would not be easy to visualize, the case of a specific range bin is shown here, whereby the initial signal in the 2D elevation and slow-time domain is reshaped to obtain a 3D cube with the additional azimuth domain to perform DOA. 

\begin{figure}[h]
 \centering
 \includegraphics[width=60mm]{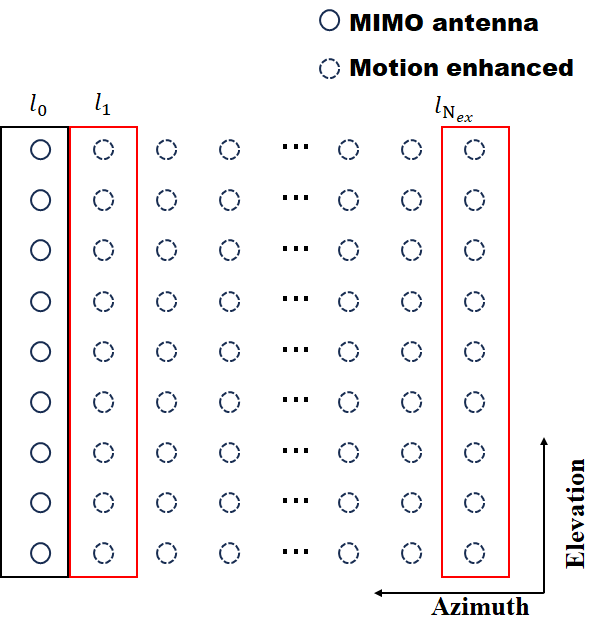}
 \caption{The complete imaging array formed by combining the MIMO virtual 2D array \& the motion-enhanced 2D arrays. The black solid line rectangle indicates the MIMO antenna positions at time $t_0$, while the dashed circles within the red rectangles indicate the motion-enhanced antennas, which are formed by the physical movement of the MIMO antenna at different time tags. The final larger array for DOA is formed by combining those together. }
 \label{fig_3}
\end{figure}

At the time $l_0$, the position of the MIMO antenna in the elevation direction would be the one in the black rectangle shown in Fig. \ref{fig_3}. 
The angular resolution will be determined by the aperture size of the array, and only the coherent channels will contribute to its improvement. After moving to a different time tag according to equation (\ref{new definition of small aperture}), the MIMO array will be physically moved to positions suitable to expand the aperture coherently. 

The maximum number $N_{m}$ of motion-enhanced snapshots that can be generated is:  
\begin{equation}
    N_{m}=\left\lfloor \frac{L_d}{\lfloor{\frac {d}{2v_yT}}\rfloor}\right\rfloor
    \label{maximum_time}
\end{equation}
where $L_d$ is the total number of chirps in one snapshot, $d$ is the distance between different receivers, $T$ is the chirp duration.

\subsubsection{Compensated steering vector}

The movement of the vehicle in the Y direction contributes to the generation of motion-enhanced aperture for azimuth sensing. However, there are two approximations in the formation of the motion-enhanced snapshots, as shown in equation (\ref{snapshot_formation}). Because of the fixed chirp duration, the position of the antenna at each starting time of the chirp will not be the same as the coherent position, leading to extra phase errors. Moreover, the movement in other, non-forward directions will introduce extra phase shifts which may lead to defocusing. Therefore, a compensated steering vector is proposed as follows. 

%The ability of an array to separate sources in space and determine their levels depends on the choice of the steering vector. It is formed by the vector representing the antenna angle of elevation and azimuth. 

The error from the approximation of coherent position within a chirp duration can be compensated through the calculated time tag in equation (\ref{new definition of small aperture}), as:

\begin{equation}
 l_e(n)= \frac {d}{2v_y}-(l_n-l_0) 
\end{equation}
where $n$ is the index in the calculation of the time tag defined in (\ref{new definition of small aperture}). 
This time delay will cause an error equivalent to an extra movement in the moving direction, thus leading to a phase error. For each motion-enhanced snapshot formed with the proposed method, the resulting phase error is written as:

\begin{equation}
\label{compensate approximation}
    w_{ea}(\theta_m,\phi_m, n)=  2v_y \times l_e(n)\sin\theta \cos\phi/\lambda
\end{equation}
where $\theta_m$ and $\phi_m$ are azimuth and elevation angle from the array boresight, respectively.

The movement in the other two, non-forward directions will lead to another phase error, which can be written as: 

\begin{equation}
\label{compensate approximation}
    w_{ev}(\theta_m,\phi_m, n)= (l_n-t_0)(v_x \cos\theta_m \cos\phi_m+v_z \sin\phi_m)/\lambda
\end{equation}

Combining the two aforementioned sources of error, a compensation factor for the original steering vector can be written as: 

\begin{equation}
    w_{er}(n,m) =w_{ea}(\theta_m,\phi_m, n)+ w_{ev}(\theta_m,\phi_m, n)
    \label{whole compensate}
\end{equation}

\subsubsection{3D high resolution imaging}
\label{3D high resolution imaging}

After FFT on the fast time domain of the signal in (\ref{fundamental_equation}), the detected targets will appear as a peak in the range profile. Then, the corresponding elements are extracted to form a new Doppler Angle Tensor (DAT) which is used for elevation DOA estimation. The angle profile of the targets can be extracted from the phase delay between different receivers. Consider a detected target at a certain range bin $r= \hat r \in [0,r_{un}]$, where $r_{un}$ is the unambiguous range, as an example. By reshaping each motion-enhanced snapshot into the third dimension, the corresponding 3D DAT $\mathbf Z(r) \in \mathbb{C}^{N_{ex} \times N_{e} \times L} $  will be obtained for joint elevation \& azimuth DOA estimation, where $N_{ex}$ is the number of the motion-enhanced snapshots generated for DOA estimation, $N_e$ is the number of antennas in elevation, and $L$ is the number of Doppler indices used for obtaining the subspace. 
In the following, all the derivations are presented for the targets in the range bin $ \hat r$ for simplicity, i.e., $\mathbf Z=\mathbf Z(\hat r)$.

% , where $n =0,1,2,...,N_{ex}$. The extended tensor $\mathbf Z$  used for 2D DOA can be formed from equation (\ref{fundamental_equation}). 

\begin{equation}
\begin{matrix}
     \mathbf Z=[\mathbf z_1&\mathbf z_1& \mathbf z_2 &\mathbf z_3 &... &\mathbf z_{N_{ex}} ]
\end{matrix}
\label{extended_tensor}
\end{equation}

% \begin{equation}
% \begin{matrix}
%  \mathbf z_b(i)=[\mathbf z(l_1)&\mathbf z(l_2)&...&\mathbf z(l_b)& ...&\mathbf z(l_{N_e})]
% \end{matrix}
% \label{extended_tensor}
% \end{equation}

where $\mathbf z_n= \hat{z}(:,l_n:l_n+L_s)$, $l_n$ is the time tag calculated in equation (\ref{new definition of small aperture}), $L_s$ is the number of slow time samples to generate the covariance matrix for DOA estimation.

Since the range is already estimated, a 2D sub-space-based algorithm can be implemented for the joint two-dimensional parameter estimation, namely the azimuth and elevation angle. The data have to be reshaped from 3-dimensional tensor form into the 2-dimensional matrix form $X \in \mathbb{C}^{(N_{ex}+1)  N_{e} \times L_d}$ by stacking azimuth and elevation dimensions together.

% \begin{equation}
% \mathbf Y(nN_e+i,l)= \mathbf Z(n,i,l)
%  % \mathbf a (\theta,\phi) \cdot \mathbf s   + \mathbf N 
% \label{signal_model_DAT}
% \end{equation}
% The DAT can be modelled as a tensor with the azimuth $n$ elevation $i$ and slow time index $l$:

The corresponding signal model will be:
\begin{equation}
    \mathbf X= \sum_{m=1}^M{\mathbf a_{\theta_m} \circ \mathbf a_{\phi_m} \mathbf S_t  +\mathbf N \in \mathbb{C}^{N_e({N_{ex}+1})}\times L_d  } 
    % &=\sum_{m=1}^M{\mathbf A(m) \mathbf S_t }+ \mathbf N \in \mathbb{C}^{N_e\times ({N_{ex}+1})}
    \label{angle_vector}
\end{equation}
where $\circ$ is the outer product, $\mathbf S_t$ is the point scatterer reflection coefficient with dimension $(N_{ex}+1)N_e\times L_d$, $\mathbf N $ is the noise component. The steering vectors for motion enhanced snapshots in azimuth and for the elevation angle are defined as:

\begin{equation}
\label{steering vector}
\begin{split}
\mathbf{a_{\theta_m}}= [1,&e^{-j2\pi \omega_a(1,m)},...,e^{-j2\pi \omega_a(p,m)}...e^{-j2\pi \omega_a(N_{ex},m)}]^{T} \\
\mathbf{a_{\phi_m} }= [1,& e^{-j2\pi \omega_e(1,m)},...,e^{-j2\pi \omega_e(q,m)}...e^{-j2\pi \omega_e(N_e-1,m)}]^{T} \\
&\omega_a(p,m)=\frac{d p \sin\theta_m \cos \phi_m }{\lambda}+ w_{er}(p,m) \\
&\omega_e(q, m)=\frac{d q \sin\phi_m }{\lambda} 
\end{split}
\end{equation}
where $w_{er}$ is the compensation factor in equation (\ref{whole compensate}), with $d=\frac{\lambda}{2}$.

The matched steering vector $\alpha(\theta_h,\phi_h) \in \mathbb {C} ^{(N_{ex}+1)N_e}$ for the azimuth angle $\theta_h$ and the elevation angle $\phi_h$ is formulated as:

\begin{equation}
\alpha(\theta_h,\phi_h)= \mathbf{a_{\theta_h}} \otimes \mathbf{a_{\phi_h}}
 \label{steering_vector}
\end{equation}
where $\otimes $ is the Kronecker product.

Sub-spaced methods such as DBF, MVDR, MUSIC can be implemented based on the defined steering vector for DOA estimation. The DBF algorithm, applicable without any prior information on the environment or target number, is selected here. The weight vector $\mathbf{w}_{DBF}$ that maximizes the output signal power of the array antenna is given by \cite{ArraySignal}:

\begin{equation}
 \mathbf{w}_{DBF}=\frac{\alpha(\theta_h,\phi_h)}{\sqrt{\alpha^{H}(\theta_h,\phi_h)\alpha(\theta_h,\phi_h)}}
 \label{DBF weight}
\end{equation}

The power of the weighted output is: 
\begin{equation}
  P_{DBF}(\theta_h,\phi_h)=E[\left| \mathbf{w}_{DBF}^{H} \mathbf{X}\right|^2]
\end{equation}
where $R_{\mathbf{X}\mathbf{X}}=E\left[\mathbf{X}^{H}\mathbf{X}\right]$ is the autocorrelation matrix of $\mathbf{X}$, and the $[\bullet]^H$ denotes the operation of conjugate transpose. 

\subsection{Summary of the proposed algorithm}
% The main steps of the proposed approach are summarized in the following steps and in 'Algorithm \ref{alg1}'.

% \noindent {\textbf{Step 1:}} Calculate an estimation of the target range.

% The range estimation of the target can be determined by the position of the point corresponding to the target after FFT. Through (\ref{dechirp}), we can obtain the frequency  $f=\mu \hat{\gamma}$ of the far-field \sinusoidal wave received by the array and associated to the target. 

\noindent {\textbf{Step 1:}} Generate motion-enhanced snapshots.

As discussed in Section \ref{Motion-enhanced snapshots}, the signal after range-FFT is analyzed. Snapshots coherent with respect to the original MIMO virtual array are selected based on the index in equation (\ref{new definition of small aperture}). These so-called motion-enhanced snapshots will form a new, larger aperture for 3D imaging.

\noindent {\textbf{Step 2:}} Compensate phase errors in the steering vector.

The approximation error in the calculation of an integer time tag and the error due to the presence of velocity components in the non-forward directions need compensation. Both errors are related to each index of the motion-enhanced snapshots, and are translated into the phase domain for further compensation. This is done via the compensation factor for the steering vector calculated as in equation (\ref{whole compensate}). 

\noindent {\textbf{Step 3:}} 3D imaging based on the new extended tensor $\mathbf{Z}$ and the compensated steering vector $\alpha(\theta_h,\phi_h)$.
\begin{figure*}[b]
    \centering
     \subfloat[]
    {
      \includegraphics[width=55mm]{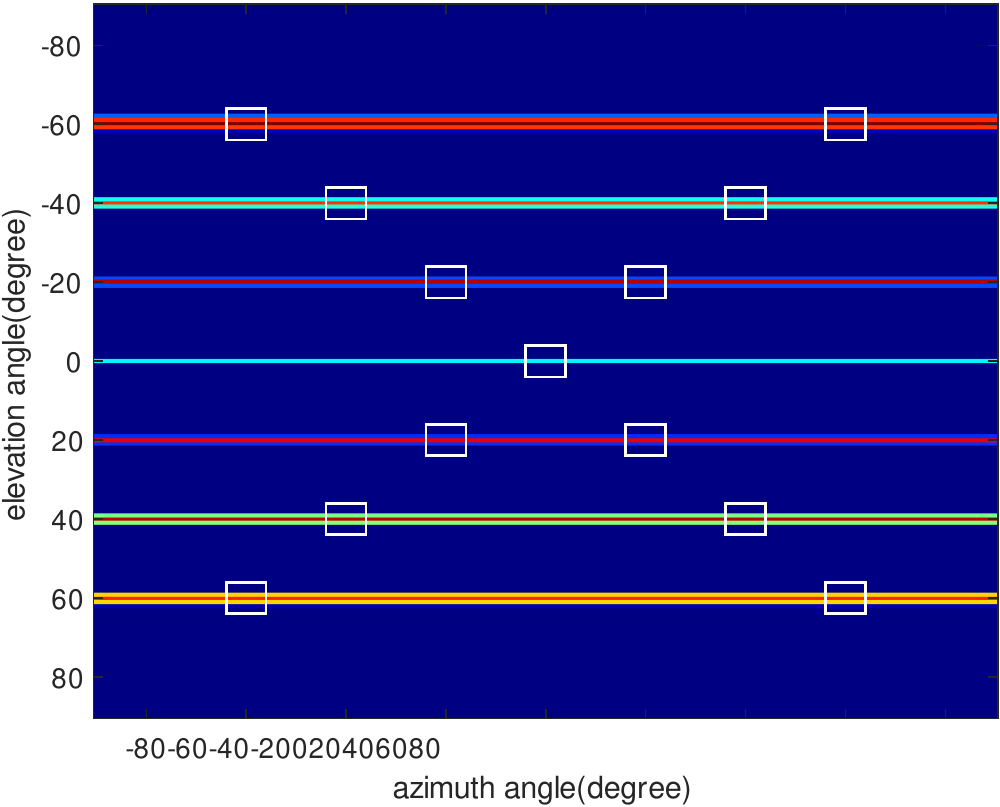}
    }
     \subfloat[]
    {
      \includegraphics[width=55mm]{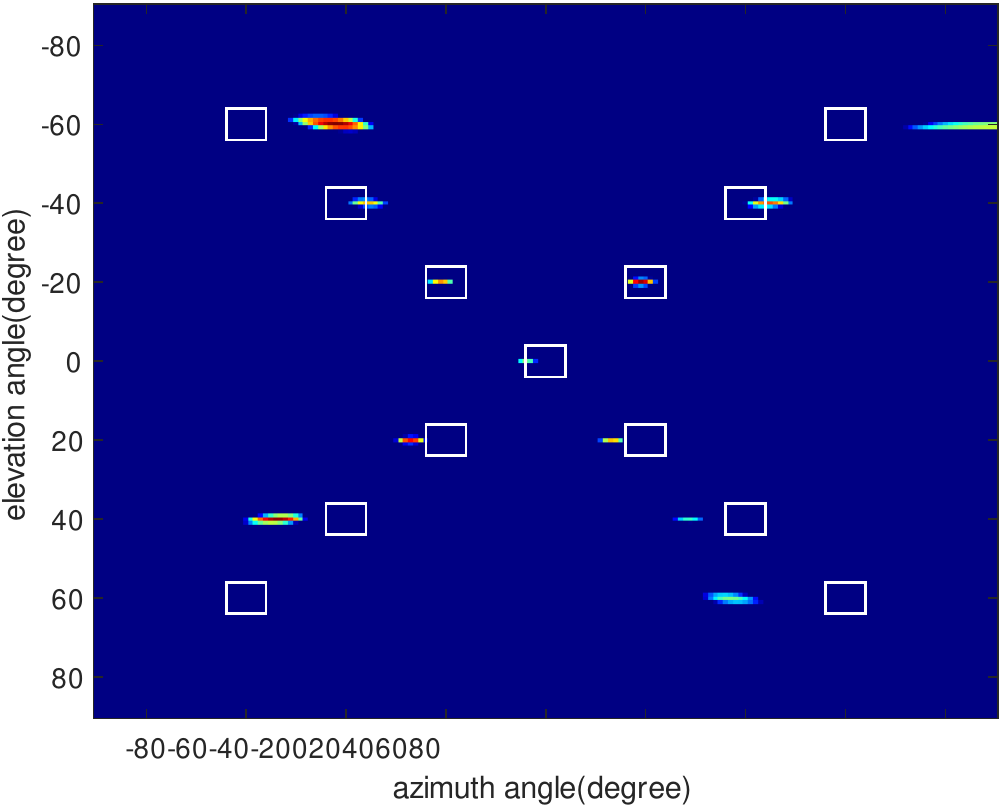}
    }
    \subfloat[]
    {
      \includegraphics[width=56mm]{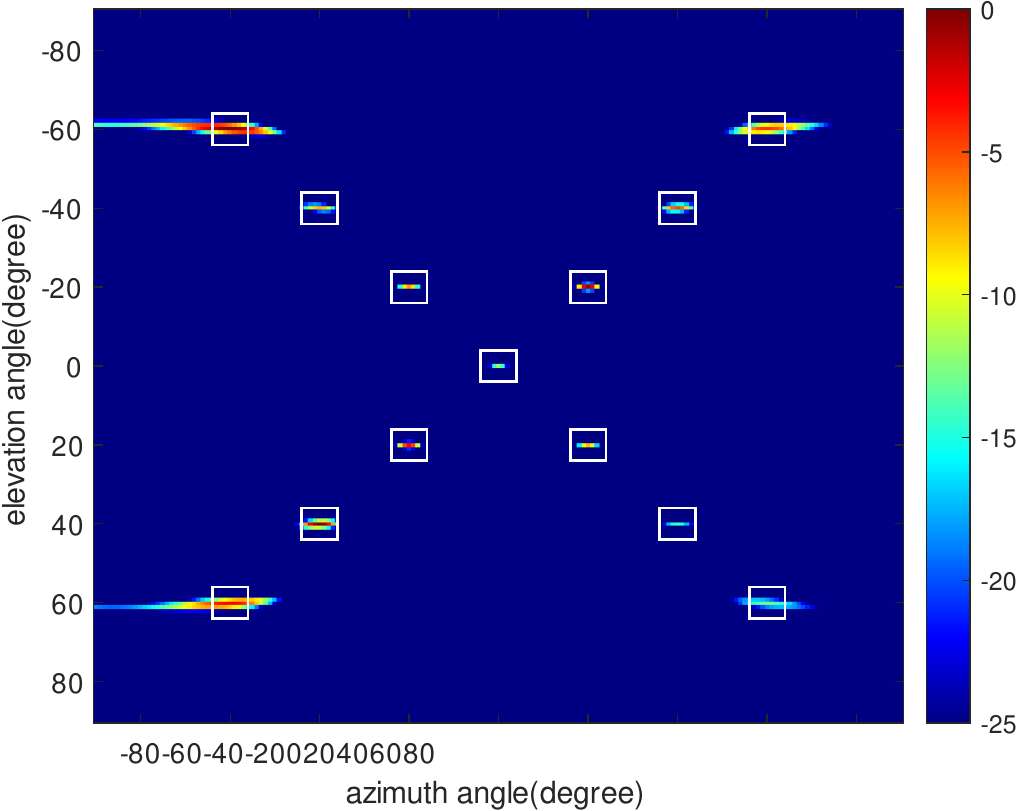}
    }
    \caption{The results of simulated 13 point targets under different methods. The sub-figures (a), (b), and (c) shows images of 1D MIMO array, proposed method without compensated steering vector, and proposed methods with compensated steering vector, respectively. The white rectangles indicate the ground truth positions of the targets.}
    \label{simulated targets}
\end{figure*}

After performing FFT along the fast-time for range estimation, the range bins with detected targets will be further processed. The tensor $\mathbf Z $ will be formed by grouping the motion-enhanced snapshots with the original antenna tensor. Then, the tensor is reshaped to a 2-dimensional matrix $\mathbf  X$ according to Section \ref{3D high resolution imaging}, and the final modified steering vector $\alpha(\theta_h,\phi_h)$ is formed with the compensation factor calculated in the previous steps.

Finally, the DBF is applied to derive the energy distribution in the azimuth-elevation domain for each range bin, thus forming the expected 3D imaging results.

\noindent
The proposed algorithm is summarized in 'Algorithm \ref{alg1}'.

\begin{algorithm}
\caption{Proposed 3D imaging algorithm}\label{alg:alg1}
\begin{algorithmic}
\STATE Perform FFT on the fast time domain and obtain the range estimation $\hat{r}$. 

\STATE Calculate the maximum number of motion-enhanced snapshots and their indexes $l_n$ as in (\ref{new definition of small aperture}).
\STATE Compute the compensated phase term for each motion-enhanced snapshot $w_{er}(n,m)$ as in (\ref{whole compensate}).
\STATE Expand the amount of snapshots for 3D imaging, generate the range-angle tensor $\mathbf{Z}$, and reshape the tensor to matrix $\mathbf  X$.
\STATE Compute the compensated steering vector $\alpha(\theta_h,\phi_h)$ as in (\ref{steering_vector}).
\STATE $\textbf{for } r $ in $[0,r_{un}]$ do
\STATE \hspace{0.5cm} $\textbf{for } \theta$ in $[-90^\circ,90^\circ]$ do
\STATE \hspace{1cm} $\textbf{for } \phi$ in $[-90^\circ,90^\circ]$ do
\STATE \hspace{1cm} $R_{\mathbf{Y}\mathbf{Y}}=E\left[\mathbf{Y}\mathbf{Y}^{H}\right]$
\STATE \hspace{1cm}  $   P_{DBF}(r,\theta,\phi)
     =\frac{\mathbf w_w^{H}((r,\theta,\phi))R_{\mathbf{Z}\mathbf{Z}}\mathbf w_w((r,\theta,\phi))}{\mathbf w_w^{H}((r,\theta,\phi))\mathbf w_w((r,\theta,\phi))}$
\STATE $ \textbf{endfor }$  
\end{algorithmic}

\label{alg1}
\end{algorithm}

\section{Results}
In this section, the effectiveness of the proposed method is showcased through the use of simulated ideal point targets, simulated complex extended targets, and experimental data.

% \subsection{Simulated and experimental results}
\subsection{Numerical simulations}
We employed a simulated 1D array in the elevation direction with 86 antenna elements, comparable to the Texas Instrument MMWCAS-RF-EVM cascade radar board AWR2243. The radar parameters are specified as follows: the starting frequency of the FMCW chirp $f_0$ is 77 GHz, the chirp bandwidth $B$ is 1 GHz, the chirp duration $T_c$ is 16 $\mu s$, the sampling rate $f_s$ is 32 Msps, and L = 512 chirps are processed in each frame. 

First, 13 ideal point targets distributed in a triangle geometry in elevation and azimuth plane are simulated. The vehicle is moving at [-1,15,2] m/s in all directions. The results from the 1D array and proposed method formed array are compared in Fig. \ref{simulated targets}. The ground truth positions of the simulated targets are marked in white rectangle boxes in the images. It can be seen that the original 1D array cannot provide any resolution ability in the azimuth direction. The proposed method without compensation for movements in non-forward directions will lead to wrongly focused positions, while the proposed method with the compensated steering vector enables to focus the targets in the right positions.
    
\begin{figure}[hbtp]
    \centering
     \subfloat[]
    {
      \includegraphics[width=30mm]{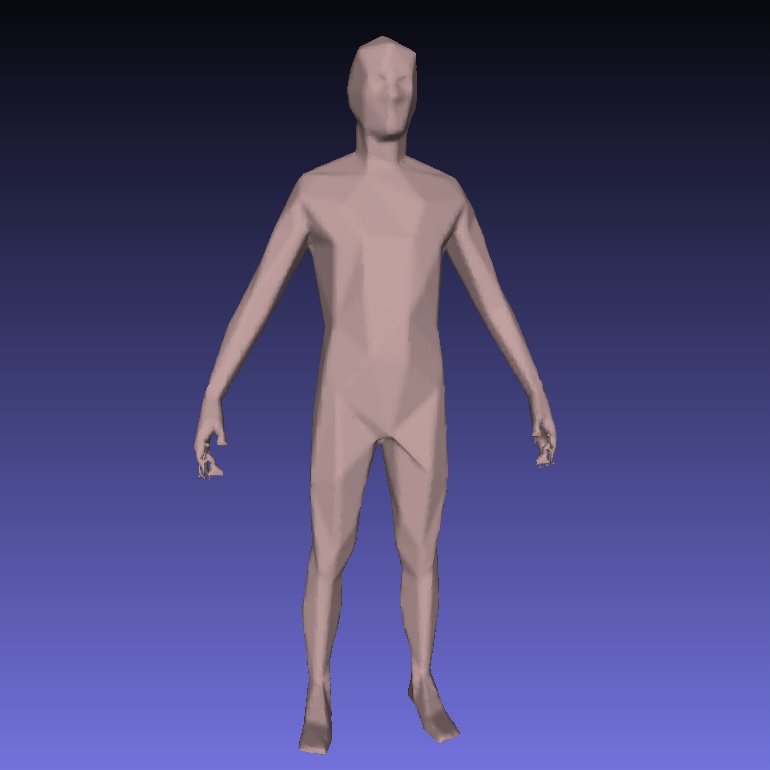}
    }
    % \hfill
     \subfloat[]
    {
      \includegraphics[width=30mm]{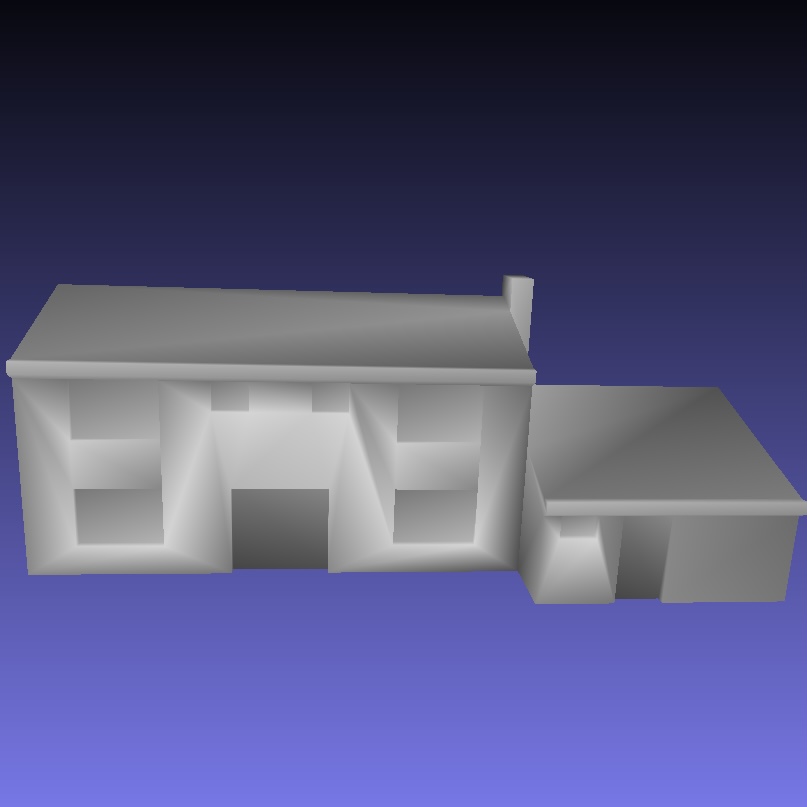}
    }
    \caption{The objects derived from CAD models for the extended targets' simulation: (a) standing pedestrian, (b) bungalow buildings.}
    \label{target_model}
\end{figure}

Simulated models of various 3D objects are then utilized as extended targets to verify the 3D imaging ability of the proposed method. These 3D objects are generated from CAD models, including a pedestrian and buildings (i.e., bungalows) as two examples of typical targets for automotive application, as shown in Fig. \ref{target_model}. All the points comprising the surface of each CAD model are used as radar scatterers within the radar field of view. To mimic realistic targets, the scatterers are resampled to ensure a uniform spatial distribution. Also, a denser sampling frequency is implemented to increase the similarity between the models and the real targets. The bungalow contains 23k scattering points, and the pedestrian contains 12k scattering points for radar simulation. It is important to note that the scatterers in the CAD models do not aim to precisely mimic electromagnetic scattering behaviour from the actual objects. Rather, they serve as a representation of the object's body shape and extent.

To simplify the subsequent analysis, certain propagation factors are not considered, such as the multipath propagation due to reflections from the road and the mutual occlusion of scatterers. These simplifications do not restrict the generality of the proposed imaging approach \cite{shengzhixu2020motion}.

The amplitudes of all scatterers are drawn from a uniform distribution, $\alpha_{\mathbf{o}} \sim \mathcal{U}(0.5,1)$, satisfying the Swerling model \uppercase\expandafter{\romannumeral3} as in \cite{yuan2022novel}. The scatterers are assumed to be isotropic and provide a constant amplitude and phase of the scattered field during the processing period, as described in \cite{andres20123d}. By employing equation (\ref{fundamental_equation}), the de-chirped signal for the scatterers representing the objects, treated as extended targets, can be simulated manually. 

\begin{figure*}[htbp]
    \centering
     \subfloat[]
    {
      \includegraphics[width=55mm]{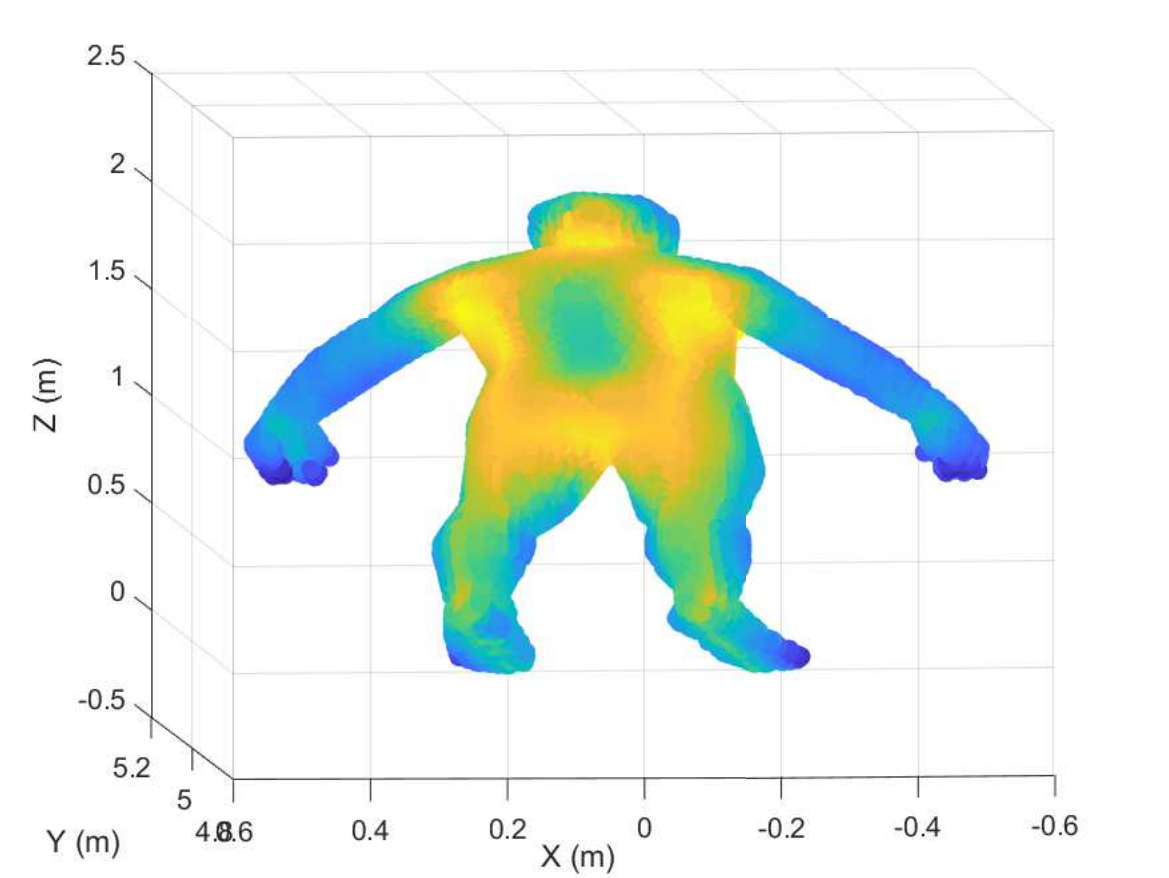}
    }
    % \hfill
     \subfloat[]
    {
      \includegraphics[width=55mm]{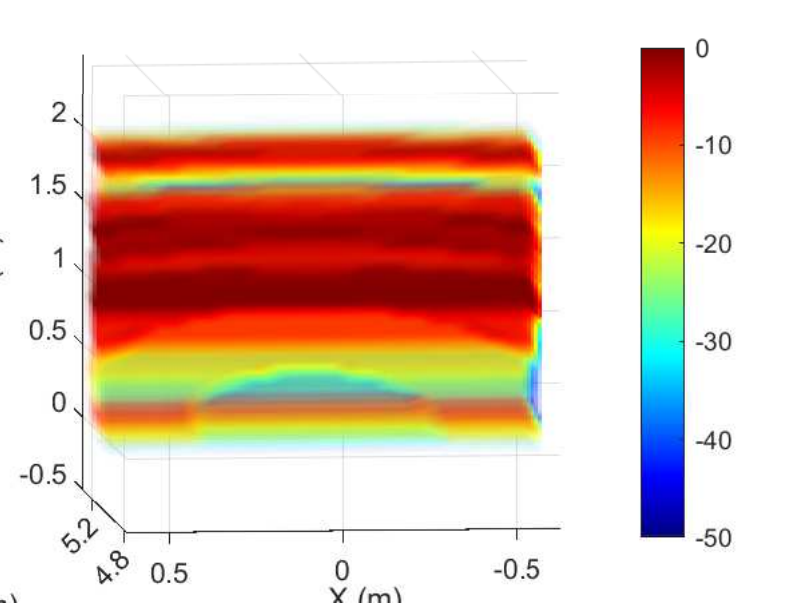}
    }
      \subfloat[]
    {
      \includegraphics[width=55mm]{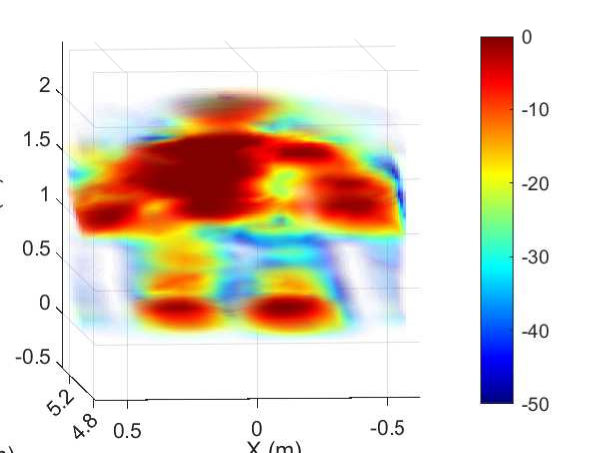}
    } 
    \caption{3D imaging results for the CAD model of a  pedestrian. The sub-figures (a), (b), and (c) show images of ground truth, conventional MIMO processing, and proposed method, respectively. }
    \label{fig: scene1}
\end{figure*}

In this simulation, a simulated 86 MIMO radar array was employed for DOA estimation, where no elements for azimuth direction and 86 elements are for elevation directions, to generate the 3D images under MIMO processing for comparison. This setting is comparable to the experimental MMWCAS-RF-EVM Texas Instrument radar board AWR2243. The radar parameters are the same as for the point targets simulation. 
For representation, the 3D imaging results with the three considered methods use mean and var quantization at first, and then are presented with the same spatial view perspective; the radar images are projected onto the Cartesian coordinates using Cubic interpolation for comparison, instead of applying another frequency algorithm to avoid creating non-existent points.

The results for the pedestrian are shown in Fig. \ref{fig: scene1}. The first column shows the original point scatterers with their respective densities, which can be considered the 'ground-truth' of these simulations, with denser scatterers located in the areas with yellow color.
The second column and third column are the 3D images generated with MIMO processing and the proposed methods, respectively. As the results in the Y and Z directions are largely impacted by the range and elevation resolution, which are the same during the processing, the results in these plane are omitted. One can easily observe that the results with proposed method obtain higher similarity with the ground truth, with more detailed shape of the target, whereas the conventional MIMO provides rather poor results in azimuth as no resolution is available in this domain. It should be noted that the slight difference in the X direction is due to the projection from the sphere coordinate to the Cartesian coordinate. 

\begin{figure*}[hbtp]
    \centering
  
     \subfloat[]
    {
      \includegraphics[width=55mm]{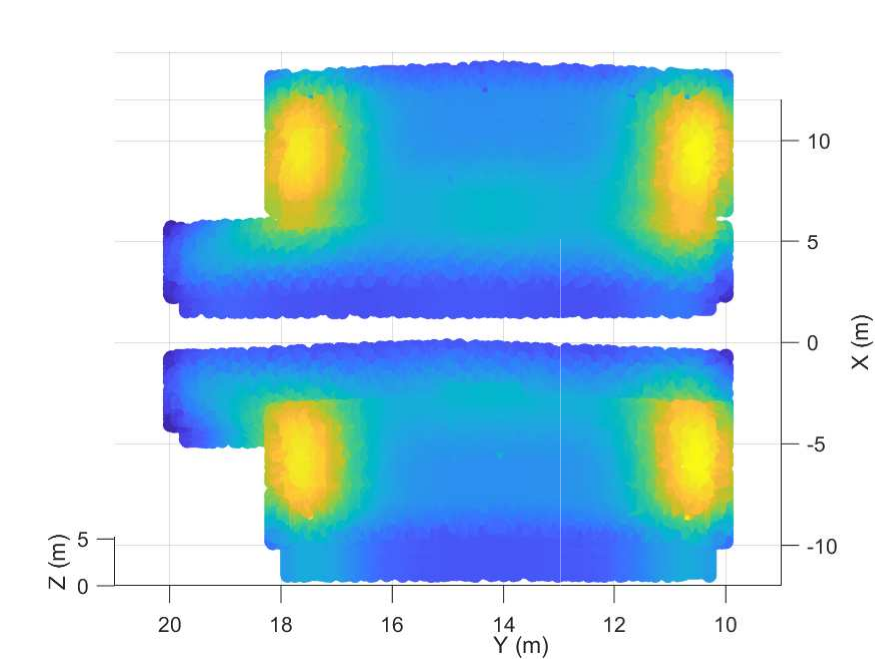}
    }
     \subfloat[]
    {
      \includegraphics[width=55mm]{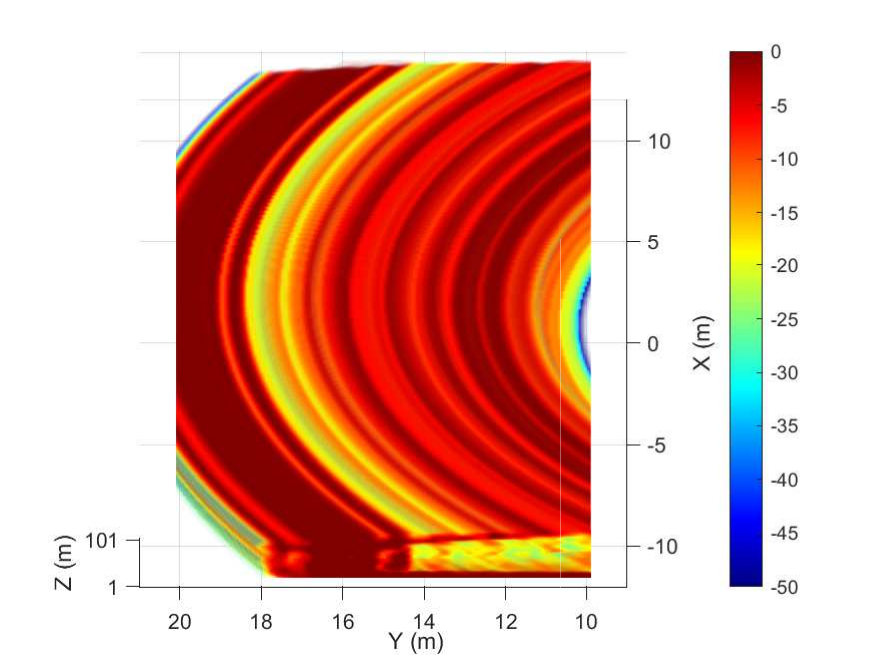}
    }
    \subfloat[]
    {
      \includegraphics[width=55mm]{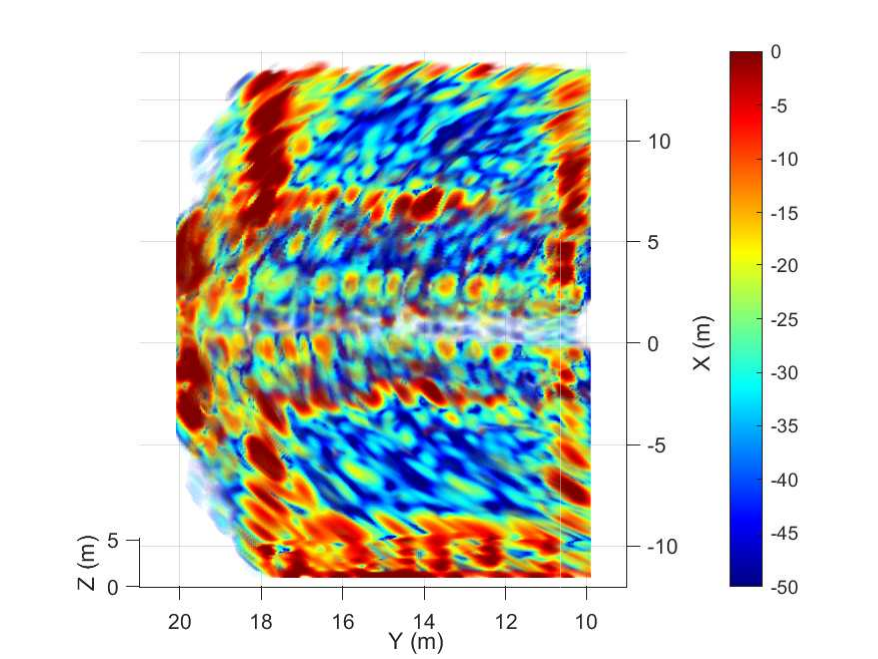}
    } \\
    \caption{3D imaging results for the CAD model of bungalow buildings. The sub-figures (a), (b), and (c) show images of ground truth, conventional MIMO processing, and proposed method, respectively. }
    \label{fig: scene2}
\end{figure*}

In a further simulation, two bungalows with a separation of 3 meters are simulated to mimic the driving scenario of small road between two buildings, common in many European cities. The results are shown in Fig. \ref{fig: scene2} with the same layout used for the pedestrian. The road between the two bungalows is visible using the proposed method as an area of lower reflectivity, whereas the result for conventional MIMO processing provides no useful information on the shape of the objects. 

These observations are based on a simple visual inspection of the images generated from the 3D objects. To further evaluate the performance, two additional quantitative evaluation metrics are introduced: voxelisation accuracy and image contrast. \cite{yuan20243DRUDAT}
% Similar to the approach in \textbf{\cite{}}, 
For voxelisation the cells in each range-elevation-azimuth tensor with intensity above +20 dB are considered to be detected to provide the target's position information. Then, the results are compared to the ground truth positions to derive the quantitative accuracy metrics shown in Table \ref{F2_score}. The number of detected cells in both ground truth and imaging results are considered as $tp$, while $fn$ is the number of cells not detected by ground truth and imaging results. $fp$ is the number of cells detected by imaging results but not ground truth, while $tn$ is the number of cells detected by ground truth but not imaging results.

\begin{equation}
\begin{split}
        &Accuracy=\frac{tp+tn}{tp+tn+fp+fn} \\
        &Precision=\frac{tp}{tp+fp} \\
        &Sensitivity=\frac{tp}{tp+fn} \\
        &Specificity=\frac{tn}{fp+tn} \\
        &AUC = \frac{Sensitivity+Specificity}{2}\\
        &F-score=\frac{2*Precision*Sensitivity}{Precision+Sensitivity}
\end{split}
\end{equation}
The proposed method is considered as a better imaging approach with its higher F-score. This achieves significant improvement in precision and accuracy. However, the AUC values for conventional MIMO are slightly better for the bungalow target, which is reasonable because the total number of detected points for the proposed method is much smaller than that for the MIMO processing, leading to a lower sensitivity results. Overall, the proposed method has better quantitative metrics than the MIMO, in addition to the qualitative higher resolution already discussed when visually comparing images.

\begin{table}[ht]
\caption{Quantitative evaluation metrics after the voxelisation for extended targets }
\small% EuMW: need this to get the 9pt text size in table cells % TODO is correct ?
\centering 
\begin{tabular}{c c c c }\midrule[1.2 pt]
Evaluation metric &Target Type & MIMO &Proposed \\  \midrule[1.2 pt]
\multirow{3}{*}{Accuracy} &Human  &22.76 &91.04 \\
&Bungalows &59.56 &92.92  \\

\multirow{3}{*}{Precision}&Human  & 0.86& 9.14 \\
&Bungalows & 2.69& 12.77 \\

% \multirow{3}{*}{Sensitivity}&Bridge & &  \\
% &Ferrari &21.03 & 13.68 \\
% &Working man &16.93 &  15.43\\
% \multirow{3}{*}{Specificity}&Bridge & &  \\
% &Ferrari &99.6 & 99.6 \\
% &Working man & 99.75& 99.75 \\
\multirow{3}{*}{AUC}&Human  & 25.73&40.01 \\
&Bungalows & 47.77&46.95 \\

\multirow{3}{*}{F-score}&Human&1.68 &12.64 \\
&Bungalows &4.78 &14.02 \\

 \midrule[1.2 pt]
\end{tabular}
\label{F2_score}
%\vspace{-\baselineskip}% remove one line of space below this table
\end{table}

The image contrast metric shows the differences in the intensity of each pixel of the image or signal, which can be used to evaluate the sidelobe suppression on SAR images \cite{yuan2020novel,berizzi1996autofocusing}. 
Each slice generated from the 3D radar images projected onto one of the 2D planes, i.e., the azimuth-elevation, the range-azimuth, and the range-elevation plane, is evaluated by calculating the mean value of each slice image contrast. The results are shown in Table \ref{image_contrast}. The proposed method obtains around twice a higher image contrast in almost every slice of the 3D imaging for all different objects. 

In this work, the \textit{dynamic range ratio}, defined as the ratio between the dynamic range of each image generated by implementing different methods, is also employed for comparison. It should be noted that the dynamic range is calculated as the difference between the maximum value and the minimum value of each image. It is shown that with the proposed method the dynamic range ratio also increases thanks to the larger amount of coherent data that are accumulated. The ratio of the bungalow equals to 7.9537; the ratio of the human is 15.4465.

\begin{table}[htbp]
\caption{Evaluation results for the image contrast metric for different extended targets in different planes}
\small% EuMW: need this to get the 9pt text size in table cells % TODO is correct ?
\centering 
\begin{tabular}{c c c c }\midrule[1.2 pt]
Target Types& Method &Image type &  Image contrast   \\  \midrule[1.2 pt]

\multirow{6}{*}{Human}&\multirow{3}{*}{MIMO} &Range-azimuth& 2.3351 \\

& & Range-elevation&  2.5077\\
& &  Azimuth-elevation&  1.2665 \\ 
& \multirow{3}{*}{Proposed} &Range-azimuth& 3.1604\\
& & Range-elevation&  3.7344\\
& &  Azimuth-elevation&  2.1182 \\\hline

\multirow{6}{*}{Bungalows}&\multirow{3}{*}{MIMO} &Range-azimuth& 1.2953 \\

& & Range-elevation &  1.3294\\
& &  Azimuth-elevation&  1.1384 \\ 
& \multirow{3}{*}{Proposed} &Range-azimuth& 3.8339\\
& & Range-elevation&  3.3602\\
& &  Azimuth-elevation&  3.6446 \\\hline

 % \midrule[1.2 pt]
\end{tabular}
\label{image_contrast}
%\vspace{-\baselineskip}% remove one line of space below this table
\end{table}

\begin{figure}[htbp]
    \centering
    \includegraphics[width=0.3\textwidth]{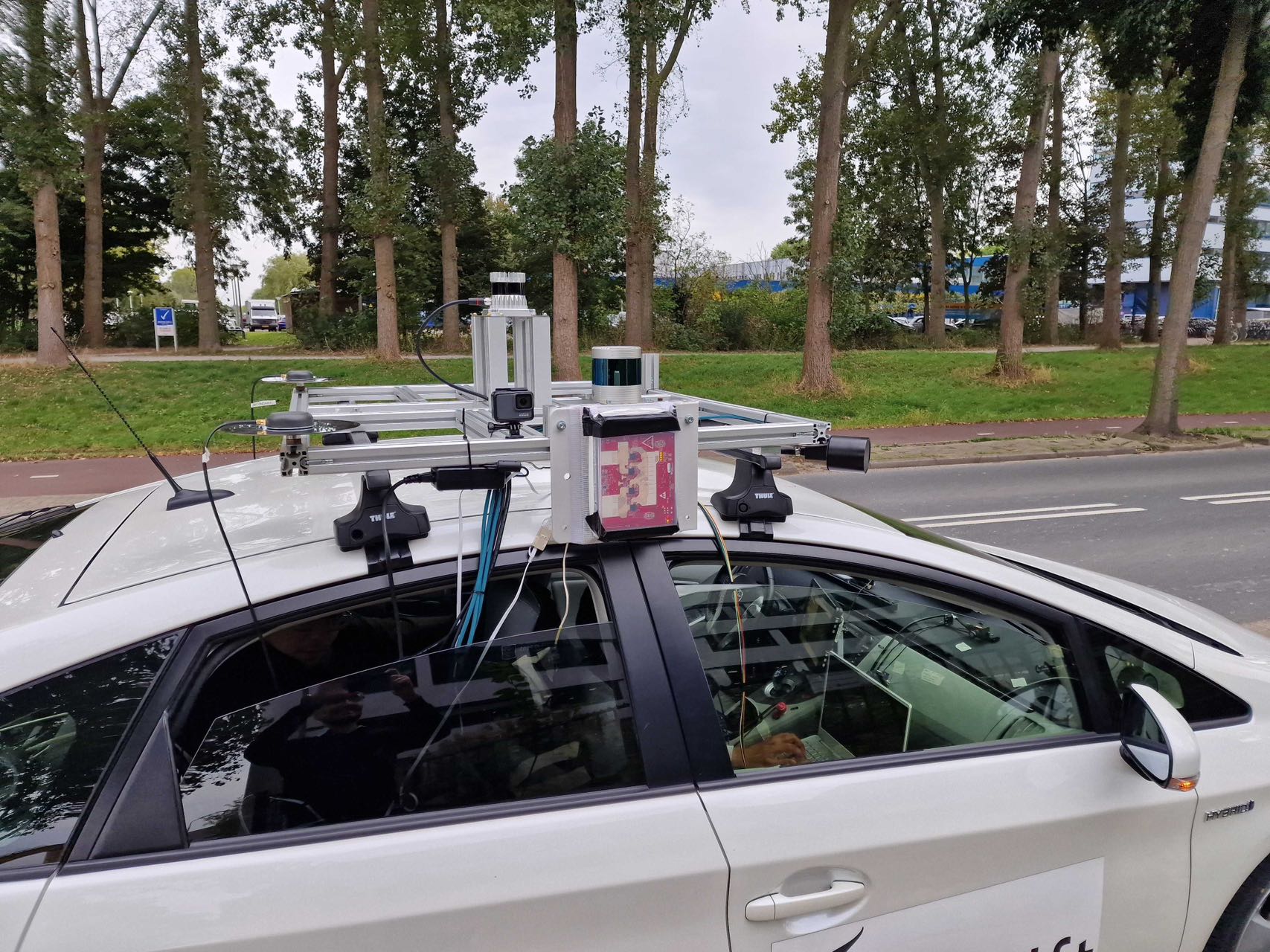}
    \caption{The experimental setup on the car with multiple sensors.}
    \label{fig:radarboard}
\end{figure}

\begin{figure}[htbp]
    \centering
    \includegraphics[width=0.35\textwidth]{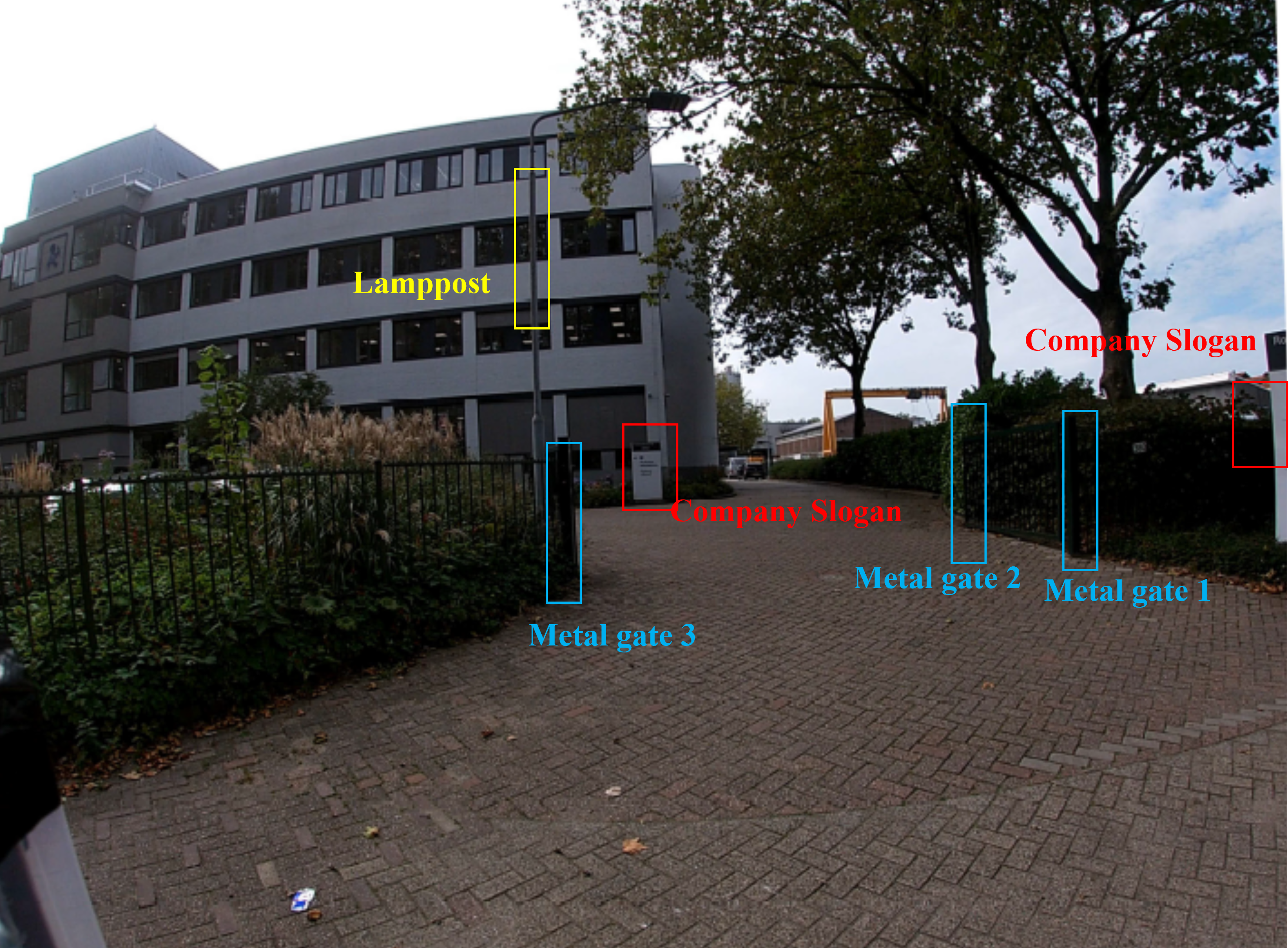}
    \caption{The optical image of the measured scene in the experimental data as captured by the GoPro camera.}
    \label{fig:SceneOfCapture}
\end{figure}

\begin{figure*}[htbp]
    \centering
     \subfloat[]
    {
      \includegraphics[width=85mm]{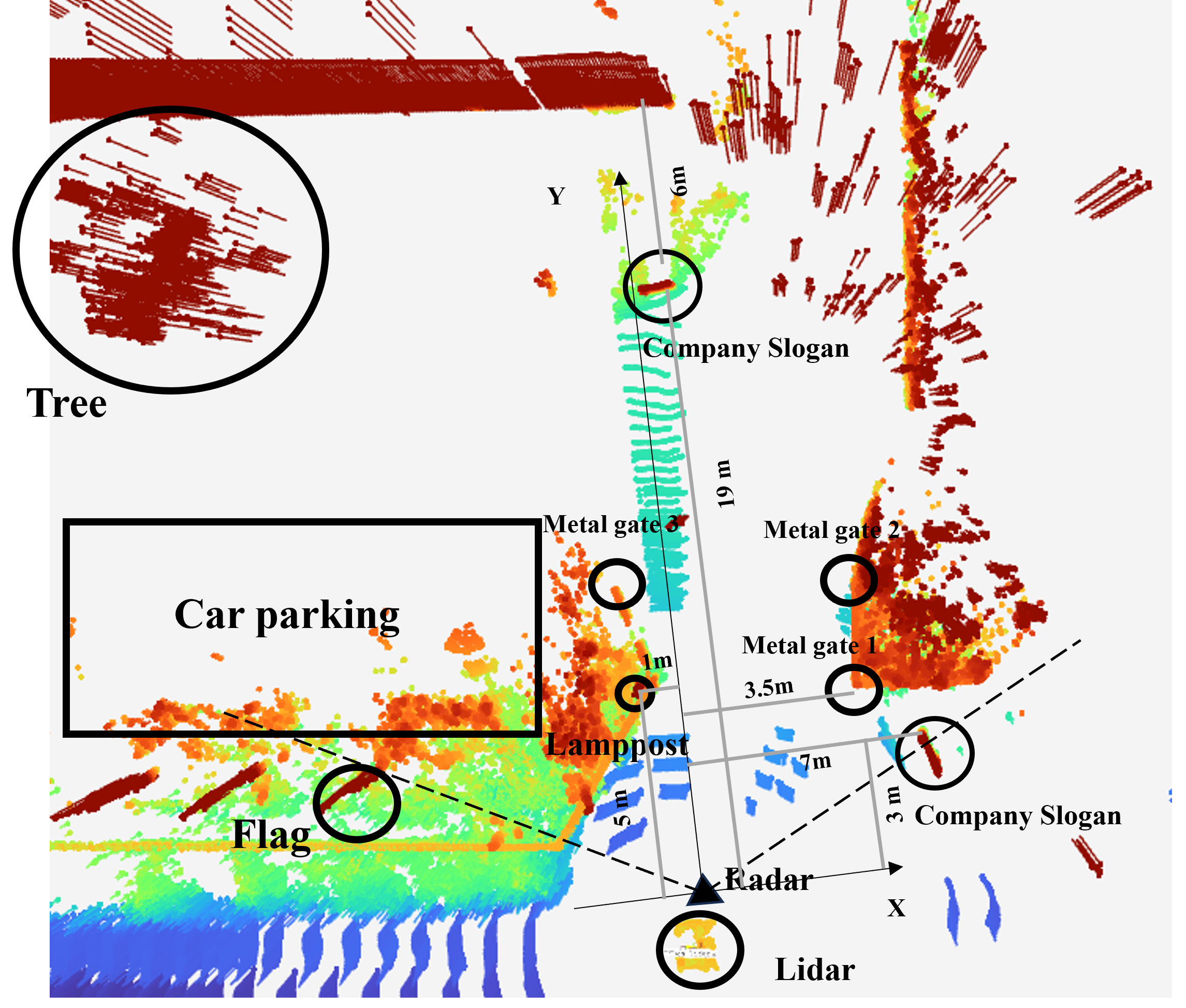}
    }
    \subfloat[]
    {
      \includegraphics[width=100mm]{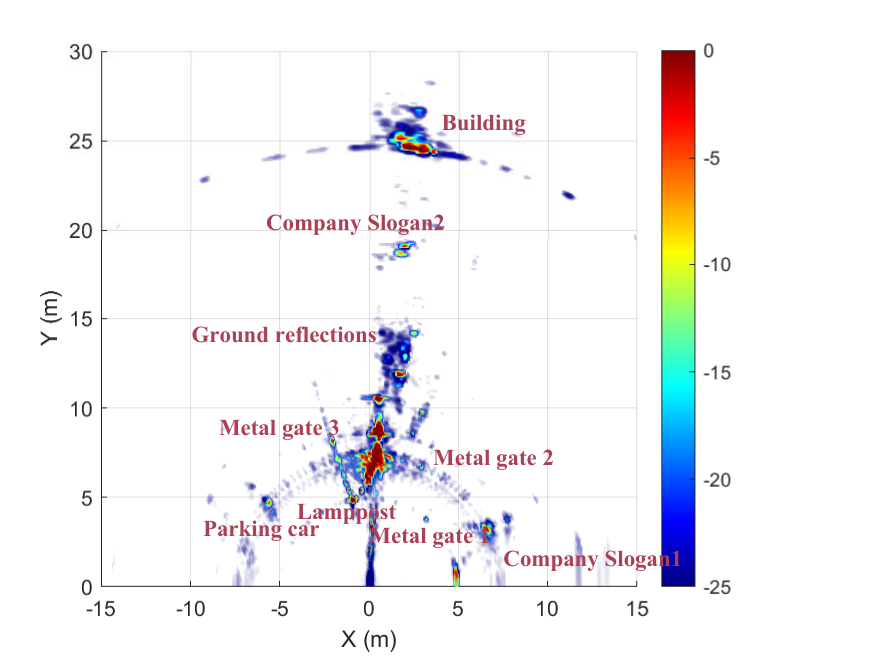}
    } \\
    \caption{Experimental results of the same scene using different sensors while driving. (a) The Lidar point cloud, where the overlaid dashed line is the radar field of view, and the added contours are the targets linked to the radar images. (b) The radar image using the proposed method with the marked targets' names.}
    \label{fig: sensors}
\end{figure*}

\begin{figure*}[hbtp]
    \centering
  
     \subfloat[]
    {
      \includegraphics[width=55mm]{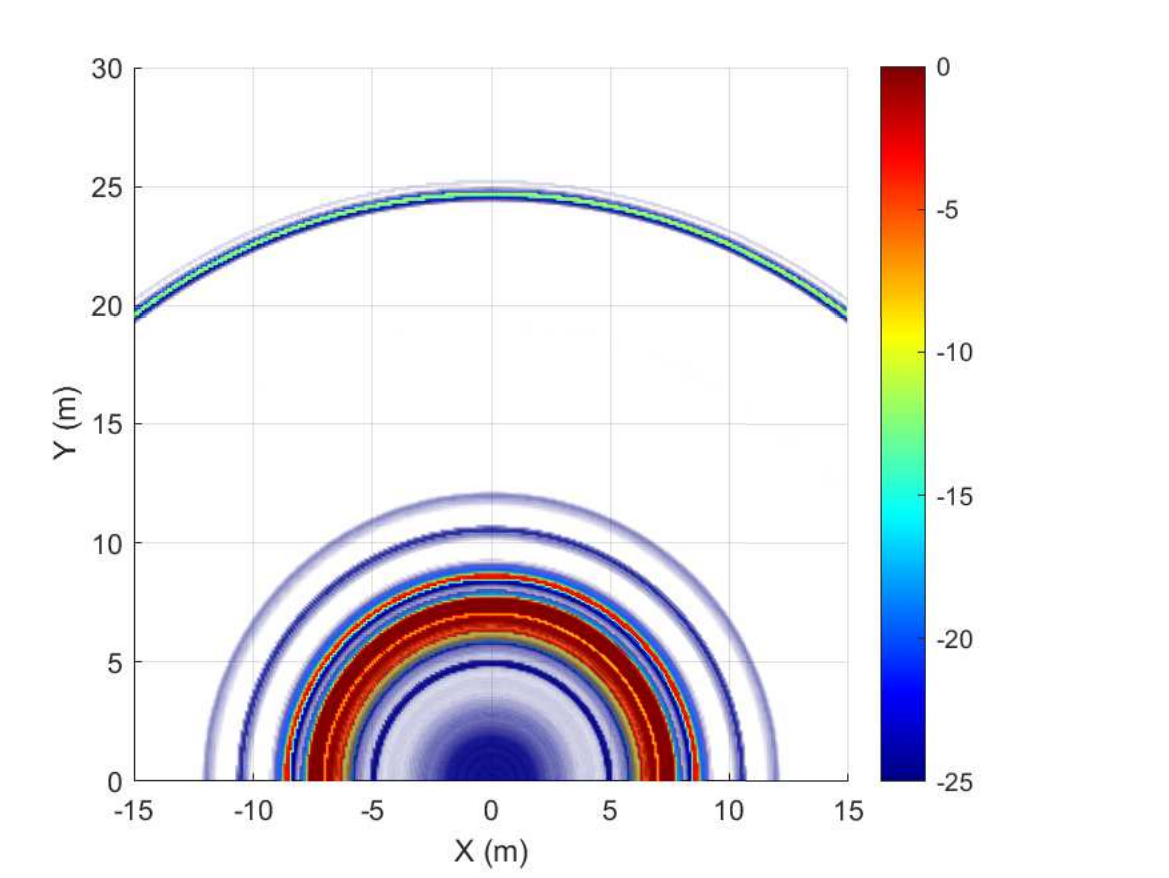}
    }
     \subfloat[]
    {
      \includegraphics[width=55mm]{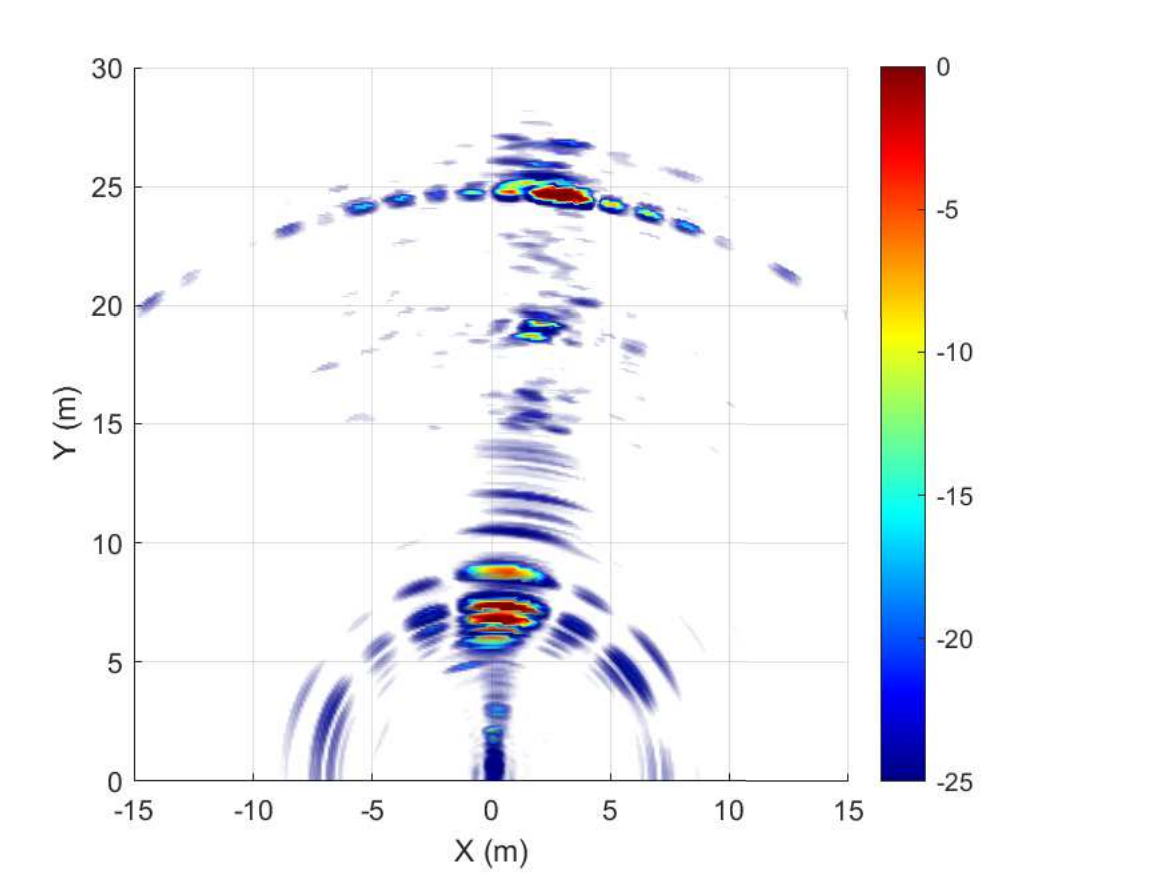}
    }
    \subfloat[]
    {
      \includegraphics[width=49mm]{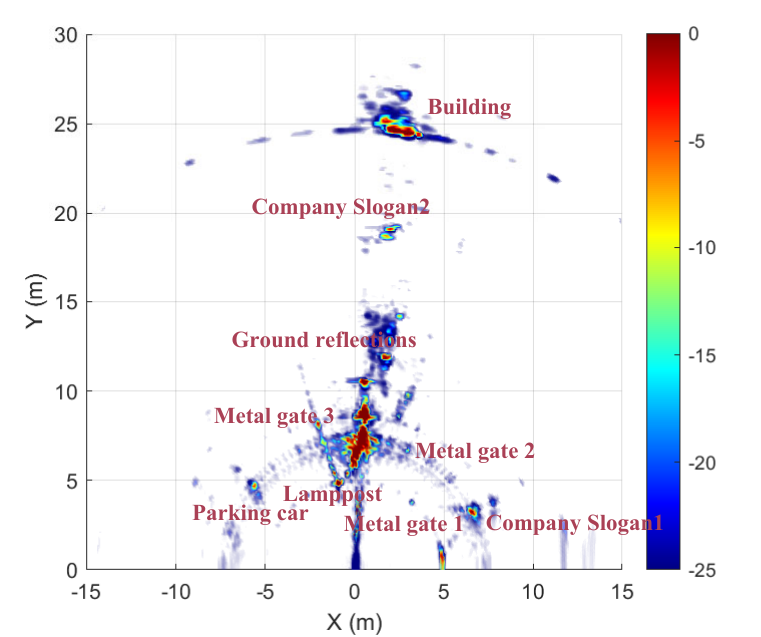}
    } \\
     \subfloat[]
    {
      \includegraphics[width=55mm]{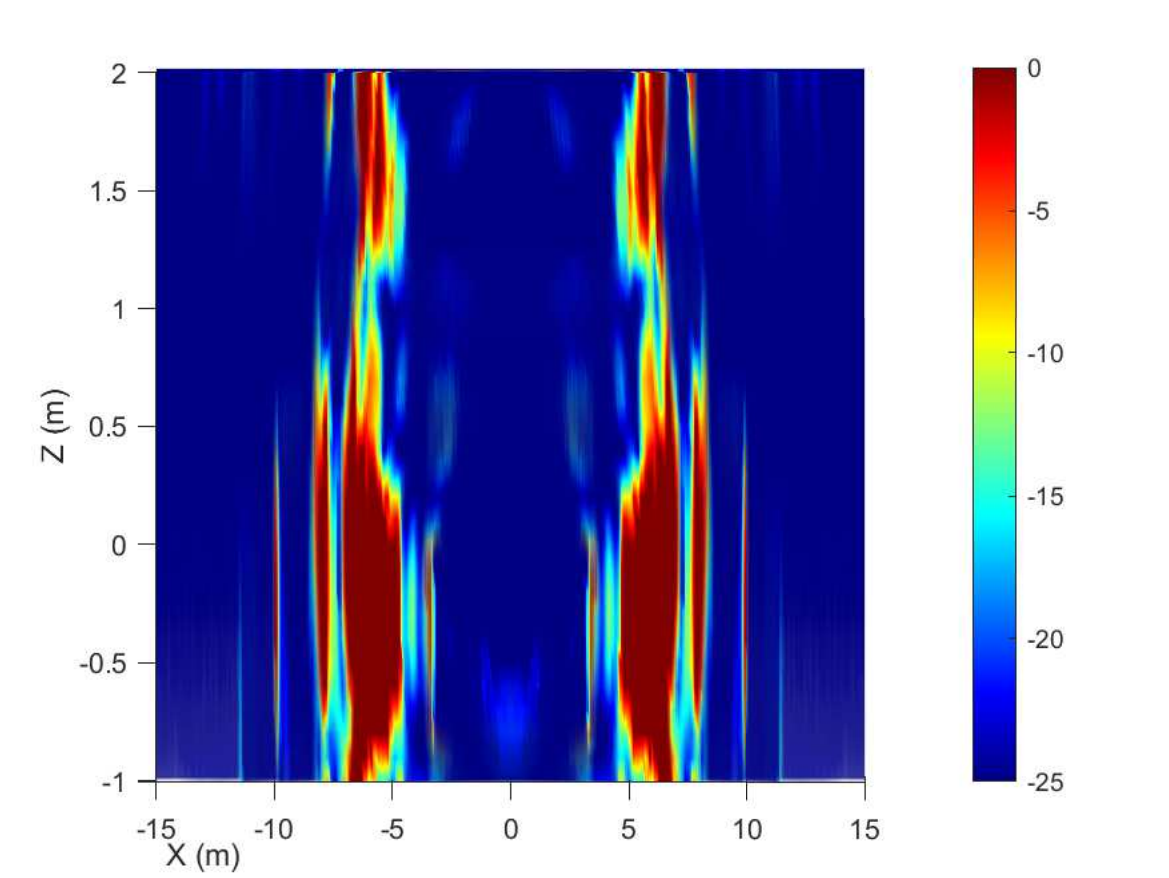}
    }
     \subfloat[]
    {
      \includegraphics[width=55mm]{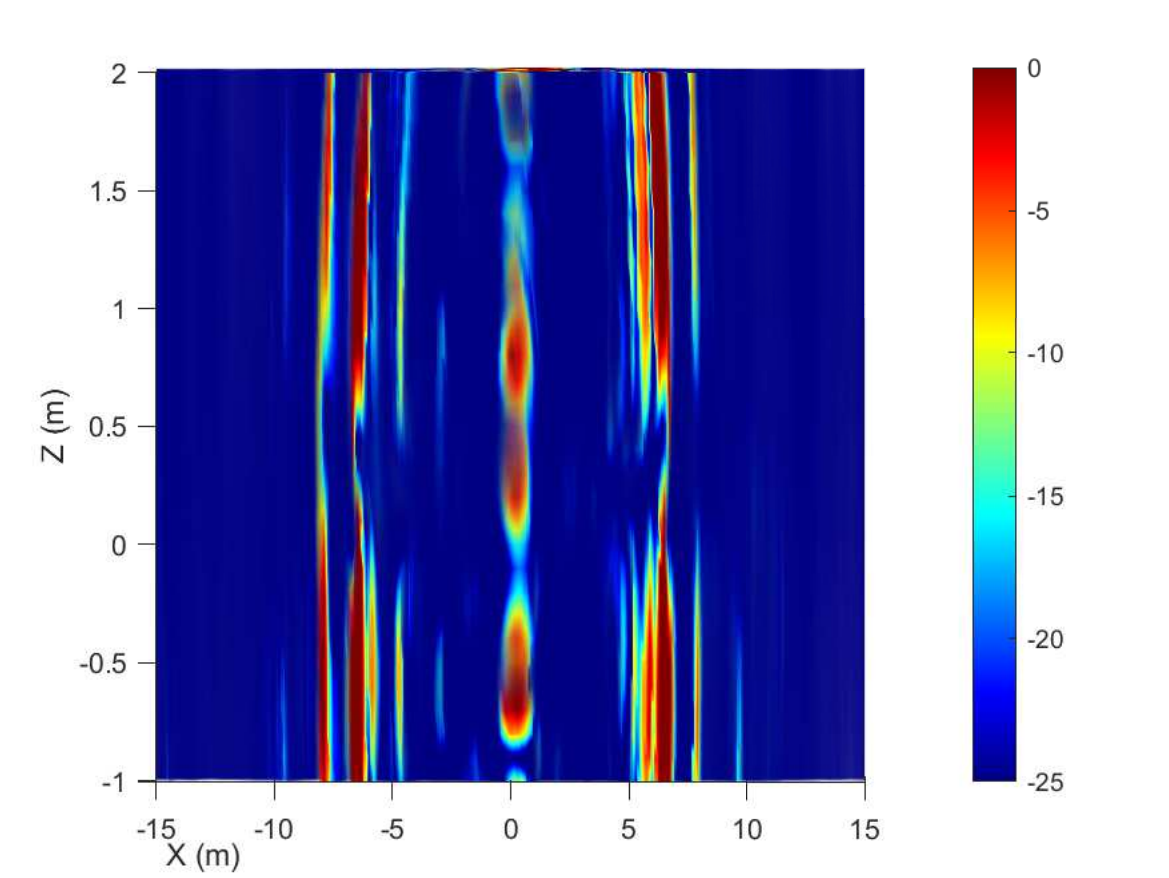}
    }
    \subfloat[]
    {
      \includegraphics[width=55mm]{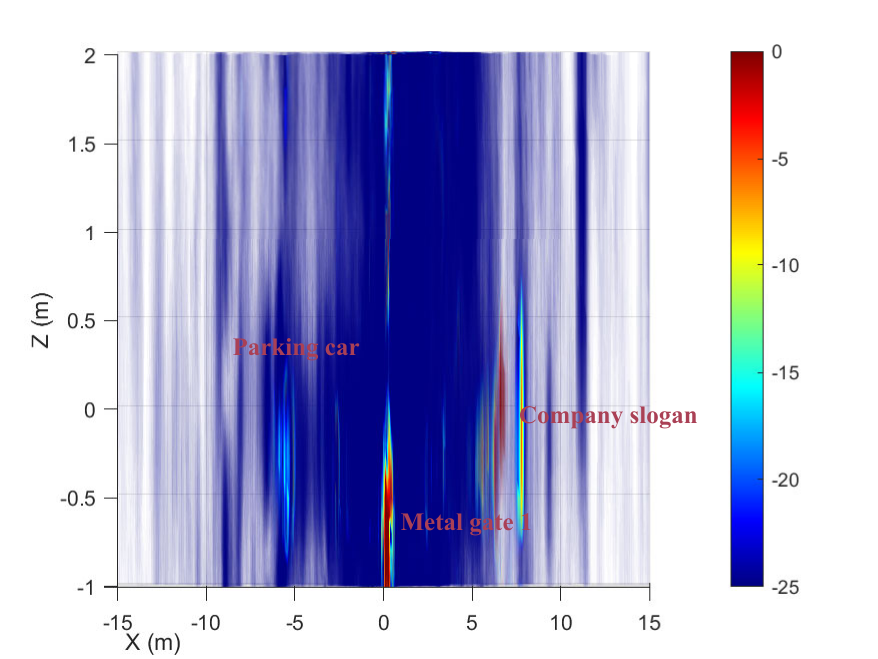}
    } \\
    
    \caption{Experimental results of the same scene while driving using different formed aperture sizes; the objects of interest in the scene were indicated in Fig. \ref{fig: sensors}. (a) The results using the original 1D array for elevation domain in BEV. (b) The results using 6 motion-enhanced snapshots in BEV. (c) The results using 64 motion-enhanced snapshots in BEV with marked targets' names. (d) The results using the original 1D array for elevation domain in forward view. (e) The results using 6 motion-enhanced snapshots in forward view. (f) The results using 64 motion-enhanced snapshots in forward view.}
    \label{fig: experimental}
\end{figure*}

\subsection{Experimental results}

The proposed approach is verified using experimental data collected with the Texas Instrument MMWCAS-RF-EVM cascade radar AWR2243, shown in Fig. \ref{fig:radarboard}. GPS, IMU, Lidar and GoPro camera are also installed on the vehicle. The radar is installed next to the GoPro, while the Lidar is on the top middle of the car. The parameters of the radar system are shown in Table \ref{parameters}. The radar is rotated in order to use its longer antenna aperture in the elevation domain, whereas DOA in the azimuth domain is performed using the method proposed in this paper. 
The radar includes 12 transmitters and 16 receivers. As the radar is working with time-division multiple access (TDMA) modulation, the movement of the car during the different transmit will cause extra error, whose correction is beyond the scope of the paper; thus, for demonstration's purposes, only the data from one transmitter are used.

\begin{table}[htbp]
\caption{Radar parameters for the experimental verification}
\small% EuMW: need this to get the 9pt text size in table cells % TODO is correct ?
\centering 
\begin{tabular}{|c|c|}\hline
\textbf{Parameters} & \textbf{Value} \\ \hline
Central Frequency (GHz) & 78.5 \\ \hline
Slope (MHz/us) & 30\\ \hline
Sampling Rate (Msps) & 6\\ \hline
Bandwidth (GHz) & 1.2 \\ \hline
Number of chirps in snapshot& 128\\ \hline
PRI(us) &75\\ \hline 
Modulation &TDMA \\ \hline 
\end{tabular}
\label{parameters}
%\vspace{-\baselineskip}% remove one line of space below this table
\end{table}

The optical image for an experimental scene considered in this work is shown in Fig. \ref{fig:SceneOfCapture}, with the corresponding image results of different sensors shown in Fig. \ref{fig: sensors}. 
The dashed line in Fig. \ref{fig: sensors}(a) is the field of view (FoV) of the radar in sub-figure (b) overlapped with the Lidar data. The distances between targets with high electromagnetic reflectivity (e.g., metal objects) in FoV and radar is measured by the Lidar and also shown in Fig. \ref{fig: sensors}(a). With the proposed method, the targets appear to be well focused on the radar image (sub-figure b) and located at the right positions. The radar image is generated with 1D elevation array with 16 antenna elements and 64 motion-enhanced snapshots for azimuth sensing. The ground reflection is high, so in the middle there appears a large energy signature. 

The results for the same scene when using different aperture sizes and in different views are shown in Fig. \ref{fig: experimental}. From the first row, it is noted that the original result do not have any azimuth sensing ability, whereas with the proposed method the azimuth sensing capability is demonstrated. From the comparison between sub-figures (b) and (c), the larger aperture formed, the better resolution obtained. In the second row, the elevation resolution is the same with the same aperture size, and the 3D results are shown in the front view starting from $y=3m$, where the company slogan side and the gate 1 are located. It should be noted that the objects of interest, such as the company slogan sign, are indicated in Fig. \ref{fig: sensors}(a).
The result in (d) do not resolve targets in azimuth, whereas in (e) and (f), the 3D imaging ability is proved. In (f) one can see that the company slogan sign is around 1m higher than the radar, while the gate 1 is 0.5m and the parking car is around 0m, matching the ground truth of the targets. In the bottom area, the ground reflection can also be observed.

\section{Discussion}
\subsection{Degrees of freedom for DOA estimation}
Vertically oriented 1D MIMO array has no degrees of freedom for DOA in the azimuth dimension. With the proposed method, 3D imaging ability is enabled by introducing motion-enhanced snapshots. The azimuth estimation is then achieved, increasing the degrees of freedom from 0 to 1.

To numerically evaluate the improvement, instead of using a 1D large MIMO array, we selected a $4+8$ MIMO radar to provide a comparison. The $+$ notation here means the usage of a L-shaped antenna, which contains 4 elements for elevation and 8 elements for azimuth estimation, with similar setting as the typical automotive radar. 
By introducing the motion-enhanced snapshots, the aperture size of the radar in the azimuth domain increases, thus increasing the number of targets whose DOA can be estimated. 
% The specifications of the radar parameters are listed as follows: the starting frequency of the FMCW chirp $f_0$ is 77 GHz, the chirp bandwidth $B$ is 1 GHz, the chirp duration $T_c$ is 16 $\mu s$, the sampling rate $f_s$ is 32 Msps, and L = 512 chirps are processed in each frame. 
The radar is moving at velocity $[-1,15,-2]m/s$. 

\begin{figure}[htbp]
    \centering
    \includegraphics[width=0.35\textwidth]{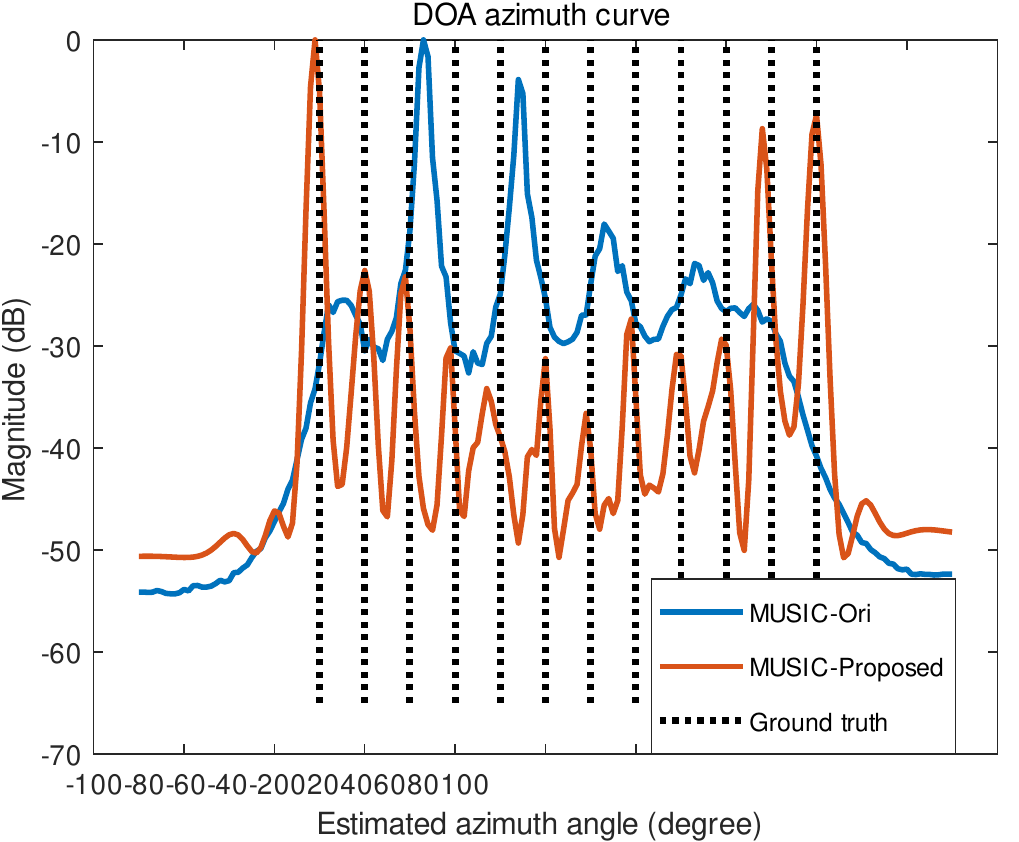}
    \caption{The simulation of 12 targets using MUSIC with application of the proposed method ('MUSIC-Proposed') and without it, i.e., with original MIMO array ('MUSIC-Ori').}
    \label{fig:enter-label}
\end{figure}

To test the performance of the proposed idea, $32*8$ motion-enhanced snapshots are selected. The MIMO antenna on the side-looking radar was located at the coordinate center, with 12 targets placed at the same range bin of 10 meters to meet the Fraunhofer distance \cite{tyler1976fraunhofer}. The targets are distributed uniformly in the azimuth dimension from $[-50\degree:60\degree]$, with a fixed elevation angle of $15 \degree$. 
The MUSIC algorithm highly relies on the rank of the sub-space matrix formed by the antenna elements, which is affected by the known rank deficiency problem. The rank is determined by the available degrees of freedom. After performing MUSIC on conventional MIMO array and the proposed array, we can easily see that the 12 targets are distinguished well using the proposed method, while this fails with MIMO, as shown in Fig. \ref{fig:enter-label}. The MUSIC algorithm can only perform well when the number of targets in a certain range bin is known and less than the number of virtual array channels. 
% The number of expanding snapshots in azimuth is influenced by the memory size of the radar, its transmission rate, and the total number of snapshots in the frame. The maximum number of motion-enhanced snapshots is given:  
% \begin{equation}
%     N_{m}=\left\lfloor \frac{L_d}{\lfloor{\frac {d}{2v_yT}}\rfloor}\right\rfloor
%     \label{maximum_time}
% \end{equation}
% where $L_d$ is the total number of chirps in one snapshot, $d$ is the distance between different receivers, $T$ is the chirp duration. 
The number of motion-enhanced snapshots is calculated as in equation (\ref{maximum_time}), thus, the degrees of freedom are increased by the same factor, i.e., $N_{m}$.

\subsection{Improvement of SNR}

% \begin{figure}[htbp]
%     \centering
%     \includegraphics[width=0.4\textwidth]{figures/azimuth_snr.pdf}
%     \caption{The simulation of 12 targets using MUSIC with and without the motion-enhanced snapshots.}
%     \label{fig:azimuth_snr}
% \end{figure}

%The signal to noise ratio (SNR) is the ratio between the signal and the noise. Higher numbers generally mean a better specification, since there's more useful information (the signal) than unwanted data (the noise). With a lower value, more noise interferes with the signal processing capabilities of your network, causing random noise and amplitude modulation. If the SNR value gets lower than one, the signal becomes unusable. This is called the 'noise floor.' Many algorithms such as matched filter are designed to gain higher SNR from the processing gain.  

The coherent combination of motion-enhanced snapshots is expected to increase useful signal power while the noise floor remains the same, hence the overall SNR is expected to be improved. Specifically, with $N$ motion-enhanced snapshots, the SNR will increase $\sqrt{N}$ times.

\begin{figure}[htbp]
    \centering
    \includegraphics[width=0.35\textwidth]{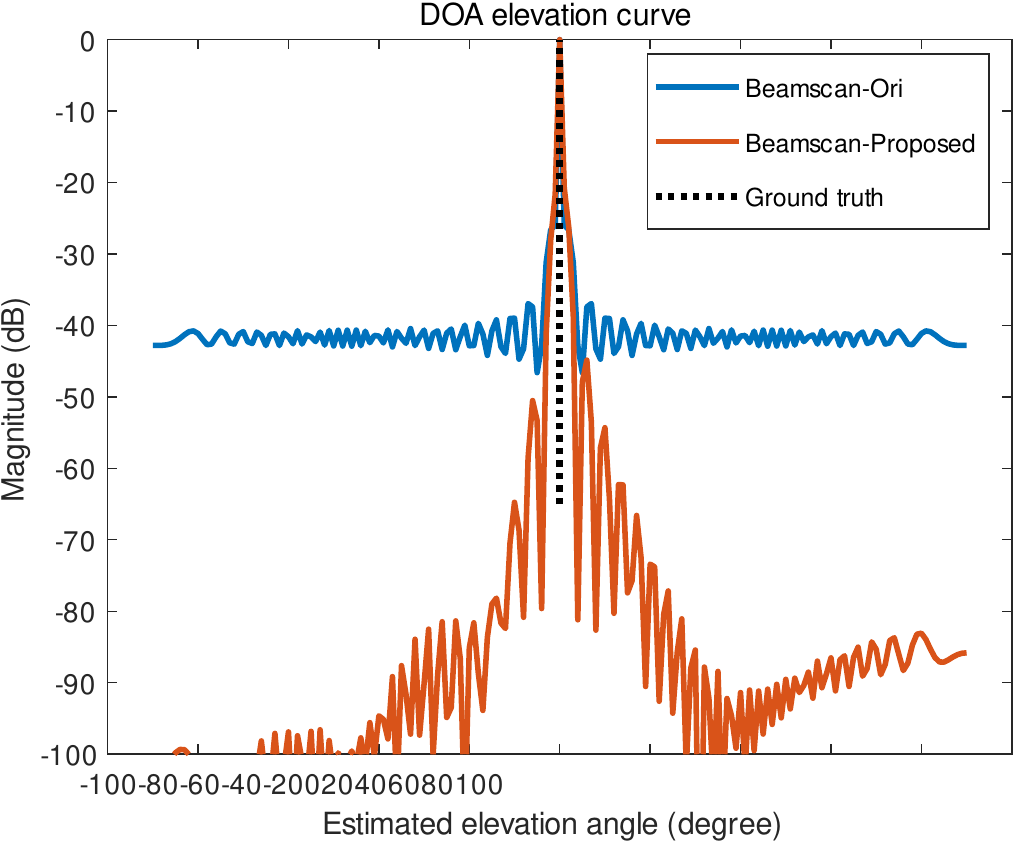}
    \caption{The simulation of SNR improvement in the elevation profile comparing the original beamscan beamformer ('Beamscan-Ori') and the beamscan with proposed motion-enhanced snapshots ('Beamscan-Proposed'). }
    \label{fig:elevation_snr}
\end{figure}

To verify the SNR improvement provided by the proposed method, an $86$ MIMO radar comparable to the Texas Instrument MMWCAS-RF-EVM radar is implemented. The radar setting and the speed are the same as in the previous section. $32*2$ motion-enhanced snapshots are selected for azimuth DOA. The target is placed at 10 m range, and azimuth \& elevation angle 0 degree. The elevation profile is shown in Fig. \ref{fig:elevation_snr}, where the SNR has increased dramatically as expected.

% \begin{table*}[ht]
% \caption{The evaluation results for the image contrast for the car model}
% \small% EuMW: need this to get the 9pt text size in table cells % TODO is correct ?
% \centering 
% \begin{tabular}{c c c }\midrule[1.2 pt]
% \textbf{Scene Type} &Image type &  Image contrast   \\  \midrule[1.2 pt]
% \multirow{3}{*}{MIMO} &Range-azimuth image& 1.836 \\

% & Range-elevation image&  3.056\\
% &  Azimuth-elevation image&  1.486 \\ \hline
% \multirow{3}{*}{Proposed} &Range-azimuth image& 3.784\\
% & Range-elevation image&  5.277\\
% &  Azimuth-elevation image&  3.632 \\\midrule[1.2 pt]
% \end{tabular}
% \label{eva_tra}
% %\vspace{-\baselineskip}% remove one line of space below this table
% \end{table*}

\subsection{Angular resolution improvement}
\label{resolution}

The spatial difference between adjacent sampling points is $\lambda/2$. If there are two targets located at $\theta+\Delta \theta$ and $\theta$, to resolve them the angle resolution $\Delta \theta$ should satisfy:

\begin{equation}
\begin{split}
  &\frac{2\pi d}{\lambda}\left(\sin(\theta+\Delta \theta)-\sin(\theta)\right) > \frac{2\pi}{N_a} \\
 &\Rightarrow  \Delta \theta_a > \frac{\lambda}{Nd\cos(\theta)}
\end{split}
\label{res_array}
\end{equation}
where $N_a$ is the number of spatial sampling point. It is noted that the resolution ability depends on the effective aperture size $Nd$. As no physical movement is present in the elevation dimension, the angular resolution will remain the same as before. The resolution improvement in the azimuth domain is instead equal to $\frac{N_{ex}+N}{N}$.

\subsection{Limitations}

The motion-enhanced snapshots necessitate that the radar physically shifts to a coherent position during slow time. If the vehicle travels too fast or too slowly, these snapshots may not be appropriate for 3D imaging. The suitability of these snapshots is contingent on the vehicle's speed, which has both maximum and minimum requirements. The maximum vehicle speed is determined by the total observation time within a single frame. In contrast, the minimum speed is associated with the chirp duration $T_c$. Combining the two requirements the vehicle velocity $V_v$ must be within:

\begin{equation}
\label{speed range}
    V_v\in [\frac{d}{ 2L_dT_{c}},\frac{d}{2T_{c}}]
\end{equation}
with typical values of $[0.1015m/s,12.98m/s]$ in our experimental settings, i.e., with one transmitter or with all the transmitters working simultaneously instead of using TDMA.
These speed requirements are crucial for the successful application of the proposed motion-enhanced snapshots for 3D imaging.

\section{Conclusion}

In this paper, a novel high-resolution 3D imaging algorithm using only 1D MIMO array in the elevation dimension is proposed for side-looking automotive radar. The compensated steering vector is proposed for addressing the 3D imaging jointly in azimuth and elevation, and errors due to non-ideal vehicle motion are compensated to further improve the performance of the method. 

The proposed method has been validated with simulated point-like and extended targets, as well as experimental data collected by our groups. Good angular resolution improvement in azimuth, and significant improvement in SNR as well as in the number of degrees of freedom for DOA is demonstrated, compared to alternative methods.
It is worth noting that the proposed approach does not need any prior information on the environment, the number, and the approximate position of targets, and it can be combined with other high resolution algorithms, such as MVDR or MUSIC. 
A potential limitations is that the movement of the targets in the scene might lead to loss of focus, and need to be addressed in future work.

\section*{Acknowledgment}

The authors are also grateful to I. Roldan, A. Palffy, and the joint team of the MS3 and Intelligent Vehicles (IV) groups of TU Delft for organizing the experimental data collection and making data available.
% \begin{thebibliography}{1}
\bibliographystyle{IEEEtran}

\bibliography{IEEEabrv,IEEEexample}

\begin{IEEEbiography}[{\includegraphics[width=1in,height=1.25in,clip,keepaspectratio]{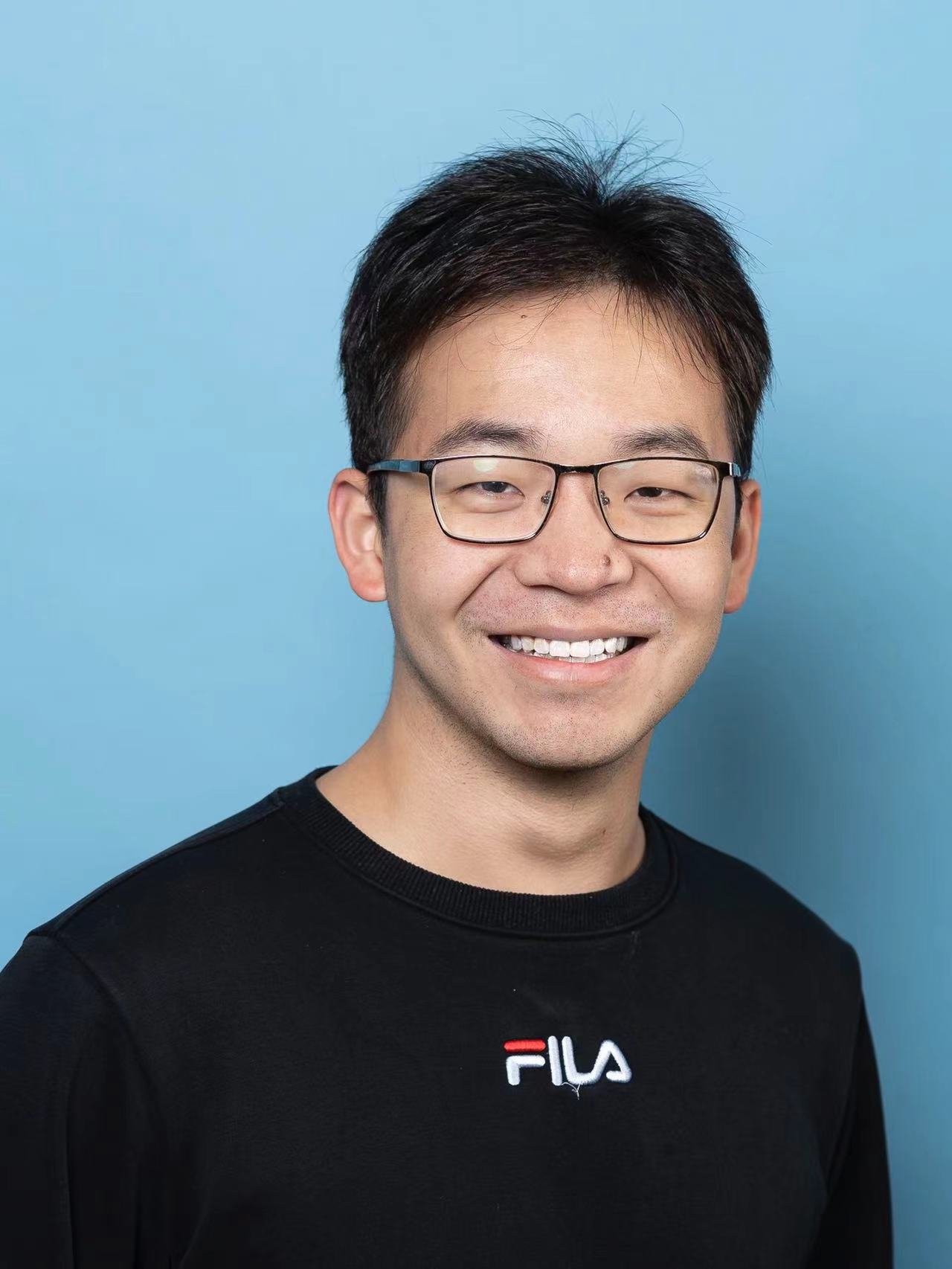}}]{Sen Yuan} was born in Shanxi province, China, in 1998. He received the Bachelor of Engineering degree in electronic information engineering from Beihang University, Beijing, China, in 2017, and the master’s degree in the specialty of signal processing from Beihang University in 2020 with Prof. C. Li and Prof. Z. Yu. He is currently pursuing the Ph.D. degree in microwave sensing signals and systems with the Delft University of Technology, Delft, The Netherlands, a section within the Department of Microelectronics.

He did an Internship with Tsinghua University, Beijing, about navigation with Dr. X. Chen in 2016. During his graduate education, he studied and became familiar with the SAR, including its satellite orbits, system design, and ground processing. In January 2021, he joined TU Delft. He works on millimeter radar signal processing in automotive applications.
\end{IEEEbiography}

\begin{IEEEbiography}[{\includegraphics[width=1in,height=1.25in,clip,keepaspectratio]{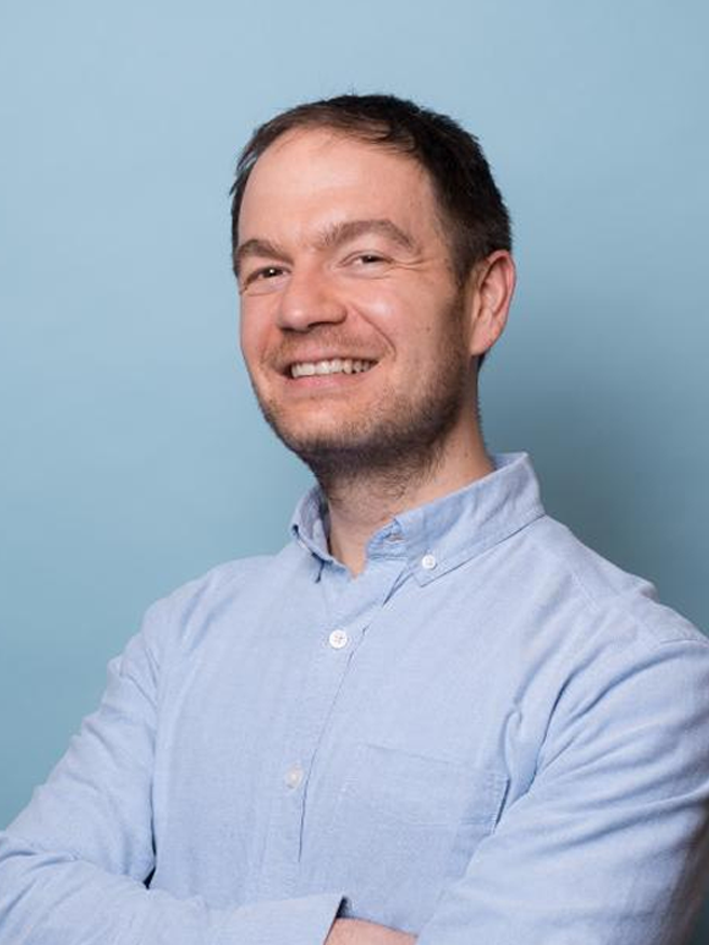}}]{Francesco Fioranelli} (M'15–SM'19) received his Laurea (BEng, cum laude) and Laurea Specialistica (MEng, cum laude) degrees in telecommunication engineering from the Università Politecnica delle Marche, Ancona, Italy, in 2007 and 2010, respectively, and the Ph.D. degree from Durham University, UK, in 2014. He is currently an Associate Professor at TU Delft in the Netherlands, and was an Assistant Professor at the University of Glasgow (2016-2019) and Research Associate at University College London (2014-2016). 

His research interests include the development of radar systems and automatic classification for human signatures analysis in healthcare and security, drones and UAVs detection and classification, automotive radar, wind farm and sea clutter. He has authored over 160 publications between book chapters, journal and conference papers, edited the books on “Micro-Doppler Radar and Its Applications” and "Radar Countermeasures for Unmanned Aerial Vehicles" published by IET-Scitech in 2020, and received four best paper awards.
\end{IEEEbiography}

\begin{IEEEbiography}[{\includegraphics[width=1in,height=1.25in,clip,keepaspectratio]{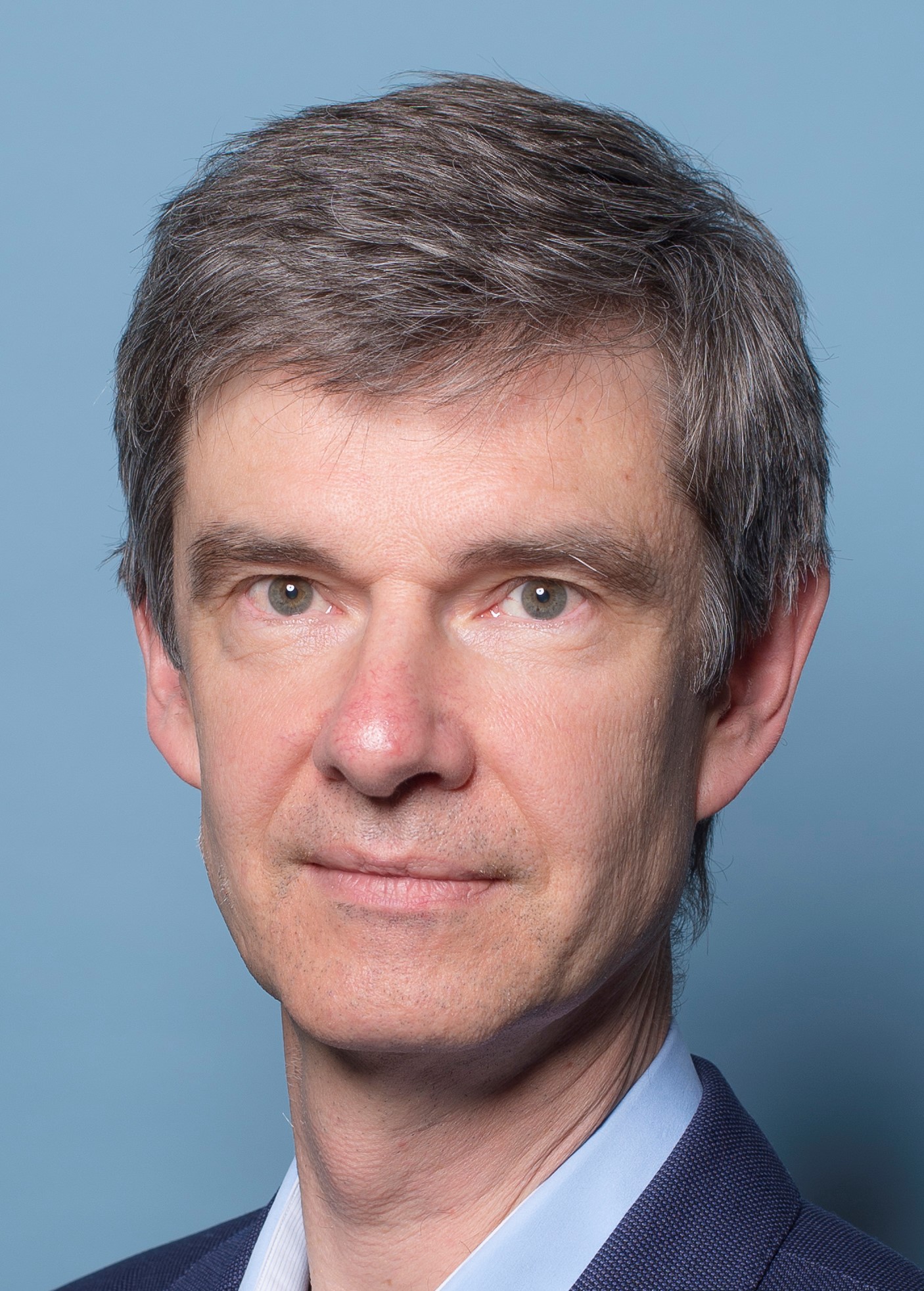}}]{Alexander G. Yarovoy} (FIEEE’ 2015) graduated from the Kharkov State University, Ukraine, in 1984 with the Diploma with honor in radiophysics and electronics. He received the Candidate Phys. \& Math. Sci. and Doctor Phys. \& Math. Sci. degrees in radiophysics in 1987 and 1994, respectively. 
In 1987 he joined the Department of Radiophysics at the Kharkov State University as a Researcher and became a Full Professor there in 1997. From September 1994 through 1996 he was with Technical University of Ilmenau, Germany as a Visiting Researcher. Since 1999 he is with the Delft University of Technology, the Netherlands. Since 2009 he leads there a chair of Microwave Sensing, Systems and Signals. His main research interests are in high-resolution radar, microwave imaging and applied electromagnetics (in particular, UWB antennas). He has authored and co-authored more than 500 scientific or technical papers, seven patents and fifteen book chapters. He is the recipient of the European Microwave Week Radar Award for the paper that best advances the state-of-the-art in radar technology in 2001 (together with L.P. Ligthart and P. van Genderen) and in 2012 (together with T. Savelyev). In 2010 together with D. Caratelli Prof. Yarovoy got the best paper award of the Applied Computational Electromagnetic Society (ACES). 
Prof. Yarovoy served as the General TPC chair of the 2020 European Microwave Week (EuMW’20), as the Chair and TPC chair of the 5th European Radar Conference (EuRAD’08), as well as the Secretary of the 1st European Radar Conference (EuRAD’04). He served also as the co-chair and TPC chair of the Xth International Conference on GPR (GPR2004). He serves as an Associated Editor of the IEEE Transaction on Radar Systems and the International Journal of Microwave and Wireless Technologies (from 2011 till 2018). In the period 2008-2017 Prof. Yarovoy served as Director of the European Microwave Association (EuMA).

\end{IEEEbiography}
\end{document}